\documentclass[apj]{emulateapj}

\usepackage{natbib}
\usepackage{xspace}

\newcommand{\hii}{{H{\scriptsize II}}\xspace}

\newcommand{\water}{H$_2$O\xspace}
\newcommand{\kms}{ km\,s$^{-1}$\xspace}
\newcommand{\miriad}{{\sc Miriad}\xspace}
\newcommand{\um}{{ $\mu{}m$}\xspace}
\newcommand{\degree}{$^{\circ}$\xspace}

\shorttitle{Accurate OH maser positions II. the Galactic Center region}
\shortauthors{Qiao et al.}

\begin{document}

\title{Accurate OH maser positions II. the Galactic Center region}

\author{Hai-Hua Qiao\altaffilmark{1, 2}, Andrew J. Walsh\altaffilmark{3}, Shari L. Breen\altaffilmark{4}, Jos\'e F. G\'omez\altaffilmark{5}, J. R. Dawson\altaffilmark{3}, Hiroshi Imai\altaffilmark{6, 7}, Simon P. Ellingsen\altaffilmark{8}, James A. Green\altaffilmark{9}, and Zhi-Qiang Shen\altaffilmark{2, 10} }
\altaffiltext{1}{National Time Service Center, Chinese Academy of Sciences, Xi'An, Shaanxi, China, 710600; qiaohh@shao.ac.cn}
\altaffiltext{2}{Shanghai Astronomical Observatory, Chinese Academy of Sciences, 80 Nandan Road, Shanghai, China, 200030} 
\altaffiltext{3}{Department of Physics and Astronomy and MQ Research Centre in Astronomy, Astrophysics and Astrophotonics, Macquarie University, NSW 2109, Australia}

\altaffiltext{4}{Sydney Institute for Astronomy (SIfA), School of Physics, University of Sydney, NSW 2006, Australia}
\altaffiltext{5}{Instituto de Astrof\'{\i}sica de Andaluc\'{\i}a, CSIC, Glorieta de la Astronom\'{\i}a s/n, E-18008, Granada, Spain} 
\altaffiltext{6}{Center for General Education, Kagoshima University, 1-21-30 Korimoto, Kagoshima 890-0065, Japan }
\altaffiltext{7}{Department of Physics and Astronomy, Graduate School of Science and Engineering, Kagoshima University, 1-21-35 Korimoto, Kagoshima 890-0065, Japan}
\altaffiltext{8}{School of Physical Sciences, Private Bag 37, University of Tasmania, Hobart 7001, TAS, Australia}
\altaffiltext{9}{CSIRO Astronomy and Space Science, Australia Telescope National Facility, PO Box 76, Epping, NSW 2121, Australia}
\altaffiltext{10}{Key Laboratory of Radio Astronomy, Chinese Academy of Sciences, China}


\begin{abstract}
We present high spatial resolution observations of ground-state OH masers, achieved using the Australia Telescope Compact Array (ATCA). These observations were conducted towards 171 pointing centres, where OH maser candidates were identified previously in the Southern Parkes Large-Area Survey in Hydroxyl (SPLASH) towards the Galactic Center region, between Galactic longitudes of $355^{\circ}$ and $5^{\circ}$ and Galactic latitudes of $-2^{\circ}$ and $+2^{\circ}$. We detect maser emission towards 162 target fields and suggest that 6 out of 9 non-detections are due to intrinsic variability. Due to the superior spatial resolution of the follow-up ATCA observations, we have identified 356 OH maser sites in the 162 of the target fields with maser detections. Almost half (161 of 356) of these maser sites have been detected for the first time in these observations. After comparing the positions of these 356 maser sites to the literature, we find that 269 (76\%) sites are associated with evolved stars (two of which are planetary nebulae), 31 (9\%) are associated with star formation, four are associated with supernova remnants and we were unable to determine the origin of the remaining 52 (15\%) sites. Unlike the pilot region (\citealt{Qie2016a}), the infrared colors of evolved star sites with symmetric maser profiles in the 1612 MHz transition do not show obvious differences compared with those of evolved star sites with asymmetric maser profiles. 
\end{abstract}

\section{Introduction}
\label{introduction}
Hydroxyl (OH) maser emission has been detected from the $^{2}{\Pi}_ {3/2}$ and $^{2}{\Pi}_ {1/2}$ rotational ladders. The most widespread OH masers are ground-state OH masers, which are from the ground rotational state $^{2}{\Pi}_ {3/2}$ ($J = 3/2$), with frequencies of 1612.231 ($F = 1 \to 2$), 1665.402 ($F = 1 \to 1$), 1667.359 ($F = 2 \to 2$) and 1720.530 MHz ($F = 2 \to 1$). Ground-state OH masers are usually associated with regions of high-mass star formation (HMSF; e.g., \citealt{Are2000}), the circumstellar envelopes of evolved giant and supergiant stars (e.g., \citealt{Nge1979}), supernova remnants (SNRs; \citealt{GR1968}), comets (\citealt{Gee1998}), or the centers of active galaxies (\citealt{Bae1982}). OH masers associated with HMSF regions are known as interstellar OH masers and are predominantly strong in the mainline transitions, i.e., 1665 and 1667 MHz. \citet{Qie2014} collated $\sim$375 HMSF ground-state OH masers from the literature, which includes all information of the interstellar ground-state OH maser sources known prior to 2014. Stellar OH masers, i.e., OH masers associated with evolved stars, often show double-horned spectral profiles at 1612 MHz (e.g., \citealt{Sea1997}, \citealt{Seb1997} and \citealt{Sea2001}) and occasionally exhibit 1665 and/or 1667 MHz OH transitions. SNRs are only associated with the 1720 MHz OH masers, which trace the interaction between SNRs and surrounding dense molecular clouds (e.g., \citealt{Fre1996}).  

Searches for ground-state OH masers have usually targeted regions likely to show maser emission, such as infrared (IR) point sources with colors indicative of high-mass protostellar objects (e.g., \citealt{Ede2007}), evolved star sources showing other species of masers, e.g., \water and/or SiO masers (e.g., \citealt{Lee1995}), SNRs (e.g., \citealt{Fre1996}) and maser sources showing other OH transitions (e.g., \citealt{Cas2004}). As introduced in \citet{Qie2016a}, several unbiased surveys, e.g., those by \citet{Cae1980}, \citet{CH1983a}, \citet{CH1983b}, \citet{CH1987}, \citet{Cas1998}, \citet{Sea1997}, \citet{Seb1997} and \citet{Sea2001}, have been conducted in certain portions of the Galactic plane. However, these surveys only selected ground-state transitions (mainline transitions at 1665 and 1667 MHz or satellite transition at 1612 MHz) of OH, thus favoured HMSF regions (\citealt{Cae1980}, \citealt{CH1983a}, \citealt{CH1983b}, \citealt{CH1987}, \citealt{Cas1998}) or evolved stars (\citealt{Sea1997}, \citealt{Seb1997}, \citealt{Sea2001}). Therefore, these previous searches suffer from biases, and the full population of OH masers are yet to be comprehensively understood.


The Southern Parkes Large-Area Survey in Hydroxyl (SPLASH) simultaneously observed all four ground-state OH transitions in an unbiased way (\citealt{Dae2014}) using the CSIRO Australia Telescope National Facility (ATNF) Parkes 64 m telescope. Compared with previous surveys, SPLASH reduced biases caused by targeted surveys or surveys only observing specific ground-state OH transitions. The survey area of SPLASH was 176 square degrees of the southern Galactic plane and Galactic Centre (\citealt{Dae2014}), between Galactic longitudes of 332\degree and 10\degree and Galactic latitudes of $-$2\degree and $+$2\degree (152 square degrees), plus an extra region around the Galactic Center, i.e., between Galactic longitudes of 358\degree and 4\degree and Galactic latitudes of $+$2\degree and $+$6\degree (24 square degrees). About 600 OH maser sites were identified by the initial SPLASH survey observations carried out with the Parkes radio telescope. These survey observations achieved a mean rms (root-mean-square) point-source sensitivity of $\sim$65 mJy (velocity resolution of 0.18\kms) in maser-optimised cubes (i.e. by enabling the `beam normalization' option in the software \textsc{GRIDZILLA} to form a dataset of Jy-scaled, point-source-optimized cubes; \citealt{Dae2014}). However, these single-dish observations were limited by spatial resolution (about 13$\arcmin$), which is insufficient to accurately identify the astrophysical objects associated with OH masers. Thus, observations with high spatial resolution are necessary to complement this survey, which is the motivation for our work.

In the SPLASH pilot region, i.e., between Galactic longitudes of 334\degree and 344\degree and Galactic latitudes of $-$2\degree and $+$2\degree (40 square degrees), 215 OH maser sites were detected in the Australia Telescope Compact Array (ATCA) observations towards 175 target fields identified in the Parkes survey observations. More than half of the 215 detected OH maser sites were discovered by these observations. The ATCA follow-up observations failed to detect OH maser emission towards 21 of the target fields (\citealt{Qie2016a}). \citet{Qie2016a} compared the location of these 215 OH masers with complementary data in the literature and was able to identify that 122 of the sites were associated with evolved stars, 64 with star forming regions, and two with SNRs. The nature of the remaining 27 OH maser remained uncertain. Only nine OH maser sites exhibit 1720 MHz OH masers, one of which is associated with a planetary nebula (PN; \citealt{Qie2016b}). This object was the second PN in which 1720 MHz OH masers were discovered. Only two other PNe are known to date to be associated with this type of emission (K3-35, \citealt{Goe2009}; Vy2-2, \citealt{Goe2016}). The 1720 MHz OH maser emission in this PN site varied between two epochs separated by 1.5 years and might trace short-lived equatorial ejections during the PN formation.

This paper is the second paper in a series presenting the ATCA maser follow-up of SPLASH, and provides accurate positions of ground-state OH masers in the Galactic Center region of SPLASH, i.e., between Galactic longitudes of 355\degree and 5\degree and Galactic latitudes of $-$2\degree and $+$2\degree. Further work outlining the polarization properties of masers is the subject of an upcoming companion paper.

\section{Observations and Data Reduction}
\label{observation}

Observations were conducted with the ATCA over five separate observing sessions on 2015 March 7 - 8, 2015 May 22, 2016 February 18 - 24, 2016 February 26 - 28 and 2016 March 4, using the array configurations 6C, 1.5C, 6B, 6B and 6B, respectively. The resultant synthesised beam typically fell in the range 13$\arcsec$ $\times$ 4.5$\arcsec$ to 20$\arcsec$ $\times$ 7$\arcsec$. The locations of the OH masers detected in the Parkes survey observations (\citealt{Dae2014}) were used as the pointing centres for the ATCA observations, and in total, 171 positions were targeted. In the cases where two masers were located within one field of view, an average of the two positions was taken as the pointing center. Each pointing center was typically observed as a series of five (or sometimes six), 4-minute cuts spread over a range of hour angles allowing for adequate uv-coverage and a total onsource observing time of at least 20 minutes.

The Compact Array Broadband Backend was configured in CFB 1M-0.5k mode (\citealt{Wie2011}) which recorded full polarization data across two 2-GHz IFs (each with 2048 channels), each with the option of finer resolution across 16 $\times$ 1-MHz ``zoom" bands, each with 2048 channels (corresponding to a channel spacing of 0.5 kHz or 0.09\kms). For these observations we concatenated zoom bands to give fine spectral resolution across broader bandwidths, mostly using seven bands for the 1612 MHz OH masers and three bands for each of the 1665, 1667 and 1720 MHz OH masers, except during observations on 2015 March 7 - 8 when four bands were used for the 1612 and 1720 MHz transitions and three were used for each of the 1665 and 1667 MHz (note that concatenated zoom bands overlap by 0.5 MHz so three zooms result in an observing bandwidth of 2 MHz). The zoom band set-up of 2015 March 7 - 8 followed our previous observations in the SPLASH pilot region. The reduced bandwidths during the 2015 March 7 - 8 observations meant that the full velocity range of two 1612 MHz OH masers detected in the Parkes spectrum was not observed in the ATCA observations (see Table~\ref{nondetection} for details) and one 1667 MHz spectrum was truncated.

Primary flux calibration was performed using the standard flux density calibrator PKS B1934$-$638 and either PKS B0823$-$500 or PKS B1934$-$638 was used for bandpass calibration. A phase calibrator was observed for 3 mins once every $\sim$20 mins and was chosen to be within 7\degree of each target. Within the longitude range of these observations only two phase calibrators were required and these were PKS B1710$-$269 and PKS B1740-517. 

During the data processing, the ten channels on each end of each spectrum were flagged, and the remaining channels were searched for emission. The velocity coverage over which maser emission was searched for in each transition was approximately $-$350 to +300\kms ($-$250 to +210\kms for the 2015 March 7 - 8 data) for 1612 MHz, $-$210 to +140\kms for 1665/7 MHz and $-$180 to +250\kms for 1720 MHz in the local standard of rest (LSR) reference frame. In two cases, the 1612 MHz OH maser emission detected with the Parkes telescope fell outside the velocity range of the ATCA observations (356.55+0.85 and 356.65+0.10 from the Parkes observations, observed with the ATCA during 2015 March 7 - 8 observations which had a narrower velocity coverage) which has prevented us from determining positions of those sources. For all other sources the ATCA setup covered the velocity range of the emission detected at each transition in the Parkes observations. The mean rms sensitivity of the ATCA observations is about 75 mJy in each 0.09\kms channel, which is about 10 mJy less sensitive than the original Parkes survey, but with a factor of two higher velocity resolution.

A detailed account of the data reduction procedure is given in \citet{Qie2016a}. Once fully calibrated image cubes were created, they were searched for maser emission by eye (along with peak intensity images, i.e. images of maximum emission along the spectral axis), which has been shown to be as accurate as an automated method (\citealt{Wae2012}, \citeyear{Wae2014}, \citealt{Qie2016a}). Once masers were identified, moment 0 images (integrated intensity images) were created over the velocity range of the identified maser spot and the maser emission was fitted with the \miriad task \textbf{imfit}, allowing precise positions for each of the identified maser spots to be derived as well as the fit error. Once the position was determined the \miriad task \textbf{uvspec} was used to extract the maser spectrum and this was used to derive the peak velocity, peak flux density, the velocity range and integrated flux density of each maser spot. Note that for the survey region of \citet{{Sje1998}} (between Galactic longitudes of $-$0.3\degree and $+$0.3\degree and Galactic latitudes of $-$0.3\degree and $+$0.3\degree), since some of our pointing centers were overlapping, we re-imaged these data cubes as a mosaic to achieve better signal to noise ratio. Seven sources (G359.716$-$0.070, G359.837$+$0.030, G359.939$-$0.052, G359.971$-$0.119, G000.060$-$0.018, G000.074$+$0.145 and G000.141$+$0.026) were obtained from this step. For these seven sources, we used the \miriad task \textbf{imspec} to extract the maser spectra from the image cubes, which gave higher signal-to-noise spectra compared to the spectra obtained from the \miriad task \textbf{uvspec}. In order to determine the velocity range we calculated the rms noise (1-$\sigma$) of the binned maser spectrum data (five channels, the velocity resolution is about 0.45\kms) and identify the velocity range showing emission greater than 3-$\sigma$. When a maser spectrum had two overlapping velocity components with a trough between them, we identify each peak as a separate spectral component if the difference in flux density between the peak of the weaker component and the lowest flux density of the trough was greater than the 1-$\sigma$ noise level. This method included most of the real emission from maser spots, but can exclude weak emission from channels where the noise may dominate. As such, the derived velocity ranges and integrated flux densities (described in Section \ref{results}) should be regarded as a guide. We note that the derived properties can be particularly difficult to accurately determine using this method when two maser spots are blended in both position and velocity.

As detailed in \citet{Qie2016a}, the absolute uncertainty in maser positions depends on several factors including the phase noise during the observations, but relative positional uncertainty is less affected, allowing us to  compare the relative positions of maser spots within one field of view. The phase noise is related to the distance between the phase calibrator and the target region, the accuracy of the known locations of the phase calibrators, the locations of the antennas, as well as the atmospheric conditions, thus it is hard to determine precisely. Therefore, in our work, we adopted a typical value of 1$\arcsec$ for the absolute positional uncertainty, which is based on previous ATCA OH maser surveys by \citet{Cas1998}. 

%

Since very few 1720 MHz OH masers were detected, we utilised this band to investigate the radio continuum properties of the sources associated with our OH maser detections. At this frequency we usually concatenated 3 (but had 4 during the 2015 March 7 - 8 observations) 1 MHz zoom bands, resulting in a total bandwidth of 2 MHz. Standard techniques for continuum data reduction using \miriad were employed. We achieved a typical rms noise of $\sim$10~mJy in the resultant images. The details of the continuum results are in Section \ref{starformation}.

\section{Results}
\subsection{Overall Summary}
\label{results}
With the ATCA we have detected OH maser emission from 162 of the 171 fields we targeted in our follow-up observations. The nine positions that we failed to detect emission towards are discussed in Section \ref{nodetections}. In total we detected 934 OH maser spots, the strongest with a peak flux density of 286 Jy and the weakest maser spot was 0.21 Jy.

As detailed in \citet{Qie2016a}, maser spots have been grouped into maser sites based on their separations. The discussion of the size of maser sites is detailed in Section \ref{size}. We have identified a total of 356 maser sites, 161 of which have been discovered by the SPLASH survey and the present follow-up observations. Of the 356 maser sites detected, 318 show 1612 MHz OH masers, 43 exhibit 1665 MHz OH masers, 57 have 1667 MHz OH masers and 12 sites show 1720 MHz OH masers. The properties of these 356 maser sites are presented in Table \ref{maintable}. Column 1 is the name for each maser spot, which is composed of the Galactic coordinates derived from their accurate positions, the frequency of the detected transition (i.e., 1612, 1665, 1667 or 1720 MHz) and a letter. For each frequency, these letters identify maser spots within the same maser site and are assigned sequentially according to their peak velocities (from low to high). Accurate positions of each maser spot (right ascension and declination) are listed in columns 2 and 3. Columns 4 and 5 give the peak flux density and integrated flux density of each maser spot. Columns 6, 7 and 8 are the peak, minimum and maximum velocities for each maser spot. Columns 9, 10 and 11 show the uncertainties in minor axis, major axis and position angle, respectively. The astrophysical identification of each maser site, which is described in Section \ref{identification}, is shown in column 12. The final column states whether the maser site is a new detection. Among these 356 maser sites, 73 sites only have one maser spot and the remaining 283 sites show more than one spot. The maser site with the largest number of spots is G356.646$-$0.321, which has a total of 20 maser spots at 1612, 1665 and 1667 MHz. 

\onecolumngrid
\LongTables
\tabletypesize{\tiny}
\tablewidth{\textwidth}
\tablecaption{\textnormal{Details of the 356 OH maser sites, derived from the ATCA observations.}\label{maintable}}
\begin{center}
\begin{deluxetable*}{lccccrrrrrrll}

\tablehead{
\colhead{Name}&\colhead{R.A.}&\colhead{Decl.}&\multicolumn{2}{c}{Flux Density}&\multicolumn{3}{c}{Velocity(\kms)}&\multicolumn{3}{c}{Relative uncertainty}&\colhead{Comments$^{a}$}&\colhead{Ref.$^{b}$}\\
\colhead{}&\colhead{(J2000)}&\colhead{(J2000)}&\colhead{Peak}&\colhead{Integrated}&\colhead{Peak}&\colhead{Min.}&
\colhead{Max.}&\colhead{Minor}&\colhead{Major}&\colhead{Position}&\colhead{}&\colhead{}\\
\colhead{}&\colhead{($^{\rm h\,m\,s}$)}&\colhead{($^{\circ~\prime~\prime\prime}$)}&\colhead{(Jy)}&\colhead{(Jy\kms)}&\colhead{}&\colhead{}
&\colhead{}&\colhead{axis}&\colhead{axis}&\colhead{angle}&\colhead{}&\colhead{}\\
\colhead{}&\colhead{}&\colhead{}&\colhead{}&\colhead{}&\colhead{}&\colhead{}&\colhead{}&\colhead{(arcsec)}&
\colhead{(arcsec)}&\colhead{($^{\circ}$)}&\colhead{}&\colhead{}\\
}
\startdata
G355.021$+$0.146-1612A&17:32:39.098&$-$33:04:15.25&0.83&1.95&13.7&12.8&17.8&0.53&1.15&$-$16.9&U&S97\\
G355.021$+$0.146-1612B&17:32:39.097&$-$33:04:15.51&1.06&1.30&20.2&18.7&21.0&0.41&0.89&$-$16.9&U&S97\\
G355.021$+$0.146-1667A&17:32:39.123&$-$33:04:15.04&0.34&0.57&13.6&12.5&16.0&0.64&1.40&$-$15.6&U&S97\\
G355.021$+$0.146-1667B&17:32:39.047&$-$33:04:15.10&0.60&3.82&24.3&16.5&29.2&0.50&1.06&$-$15.6&U&S97\\
\\
G355.110$-$1.697-1612A&17:40:20.167&$-$33:59:14.45&4.23&12.75&12.5&11.9&22.8&0.18&0.35&$-$17.3&ES-SEV&S97\\
G355.110$-$1.697-1612B&17:40:20.168&$-$33:59:14.42&2.41&9.45&37.2&23.7&38.2&0.21&0.43&$-$17.3&ES-SEV&S97\\
G355.110$-$1.697-1612C&17:40:20.166&$-$33:59:14.41&1.57&3.27&39.3&38.7&41.9&0.28&0.57&$-$17.3&ES-SEV&S97\\
G355.110$-$1.697-1667A&17:40:20.156&$-$33:59:14.54&0.47&0.24&12.5&12.1&13.0&0.87&1.80&$-$17.3&ES-SEV&S97\\
\enddata

\tabletypesize{\footnotesize}
\tablenotetext{}{\textbf{Notes.} The first column lists the name of each maser spot, which is obtained from the Galactic coordinates of the accurate positions, followed by the frequency of the detected OH transition and a letter to denote the sequence of maser spots in the spectrum. The second and third columns are the equatorial coordinates of each maser spot. The fourth and fifth columns list the peak flux density and the integrated flux density. The sixth, seventh and eighth are peak, minimum and maximum velocities, respectively. The ninth, tenth and eleventh columns are minor axis uncertainty, major axis uncertainty and position angle of each maser spot. The twelfth column is the astrophysical identification of each maser site, followed by the reason for the identification (see Section \ref{identification}). The thirteenth column denotes new maser sites with `N' or lists the reference for previously detected maser sites.}

\tablenotetext{a}{Maser sites with unknown associations are listed as U. The rest of this column is formatted as `Assignment-Reason', where `Assignment' can be SF -- star formation, ES -- evolved star, PN -- planetary nebula or SN -- supernova remnants. `Reason' can be MMB -- association with a class II methanol maser site which indicates the presence of a high-mass star \citet{Bre2013}, based on the Methanol Multibeam survey (\citealt{Cae2010}); HOP -- based on HOPS identification \citep{Wae2014}; VIS -- based on visual check on the GLIMPSE or WISE image together with spectral features; or based on identifications found through a search of SIMBAD: LEB --  \citet{Lee2003}; SAM -- \citet{Sae2003}; TER -- \citet{TO1988}; SC1 -- \citet{Sce2003}; SC2 -- \citet{Sce2000}; SOS -- \citet{Soe2013}; LIN -- \citet{Lie1992}; TEL -- \citet{Tee1991}; ZIJ -- \citet{Zie1989}; YUS -- \citet{Yue2009}; BOW -- \citet{Boe1978}; VAN -- \citet{Vae2001}; BUR -- \citet{Bue1990}; HAB -- \citet{Hae1983}; SEV -- \citet{Sea1997}; GOM -- \citet{Goe2008}; KIM -- \citet{Kie2004}; MAT -- \citet{Mae2005a}; MES -- \citet{Mee2002}; SLO -- \citet{Sle2010}; SUA -- \citet{Sue2006}; TAF -- \citet{Tae2009}; ARG -- \citet{Are2000}; WYZ -- \citet{WY2002}; YU1 -- \citet{Yue1999}}

\tablenotetext{b}{Maser sites discovered by the SPLASH survey are listed as `N' for new. The remaining maser sites are tabulated with references to previous observations: S97 (\citealt{Sea1997}), C98 (\citealt{Cas1998}), C81 (\citealt{Cae1981}), F96 (\citealt{Fre1996}), Y95 (\citealt{Yue1995}), L92 (\citealt{Lie1992}), T91 (\citealt{Tee1991}), Z89 (\citealt{Zie1989}), Y96 (\citealt{Yue1996}), Y99 (\citealt{Yue1999}), A00 (\citealt{Are2000}), H83 (\citealt{Hae1983}), M98 (\citealt{Mee1998}), B94 (\citealt{Ble1994}), T09 (\citealt{Tae2009}) and C83 (\citealt{CH1983a}).}

\tablenotetext{}{(This table is available in its entirety in machine-readable form.)}

\end{deluxetable*}
\end{center}

\twocolumngrid

\begin{figure*}
\includegraphics[width=0.85\textwidth]{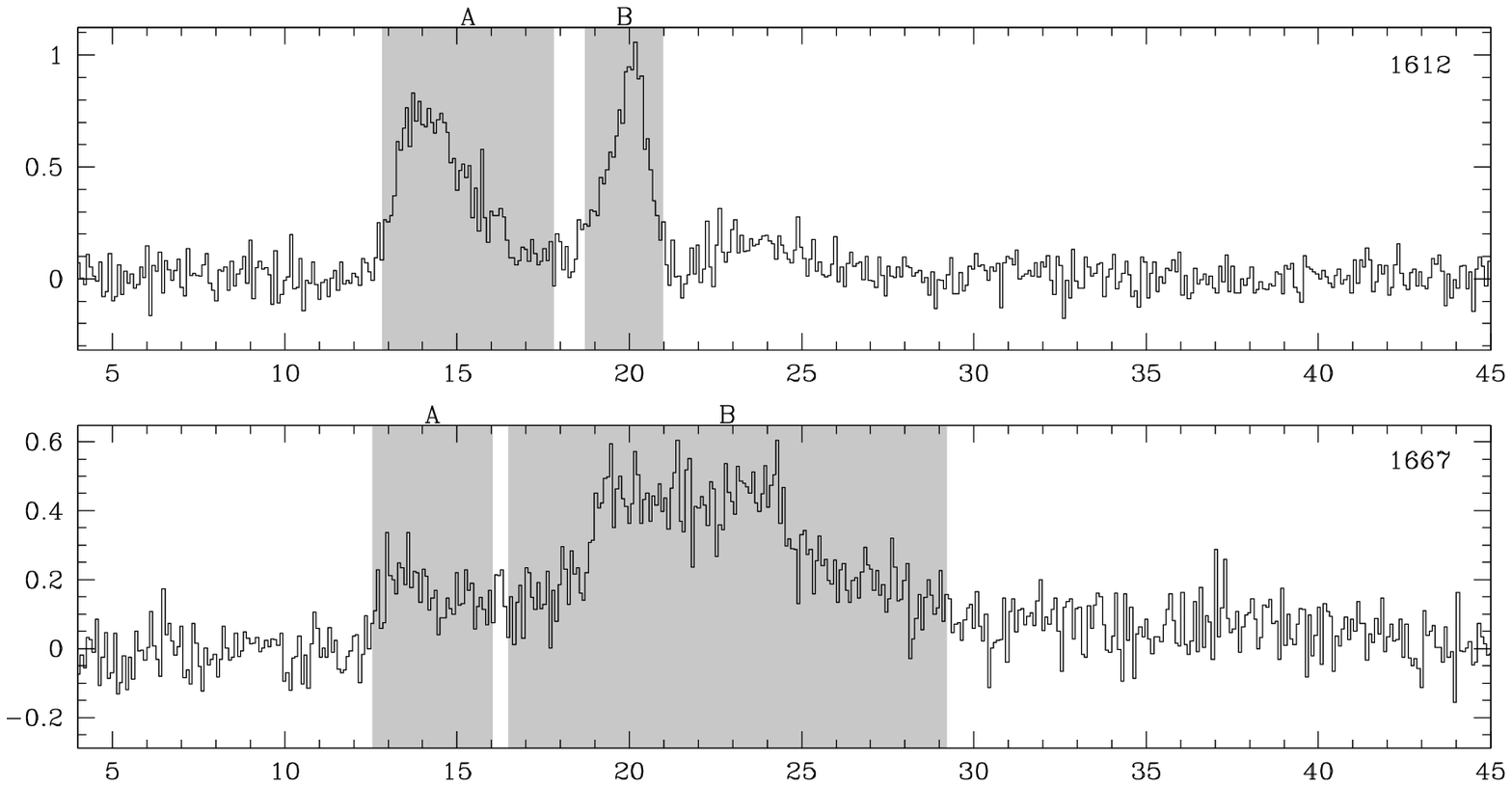}
\includegraphics[width=0.85\textwidth]{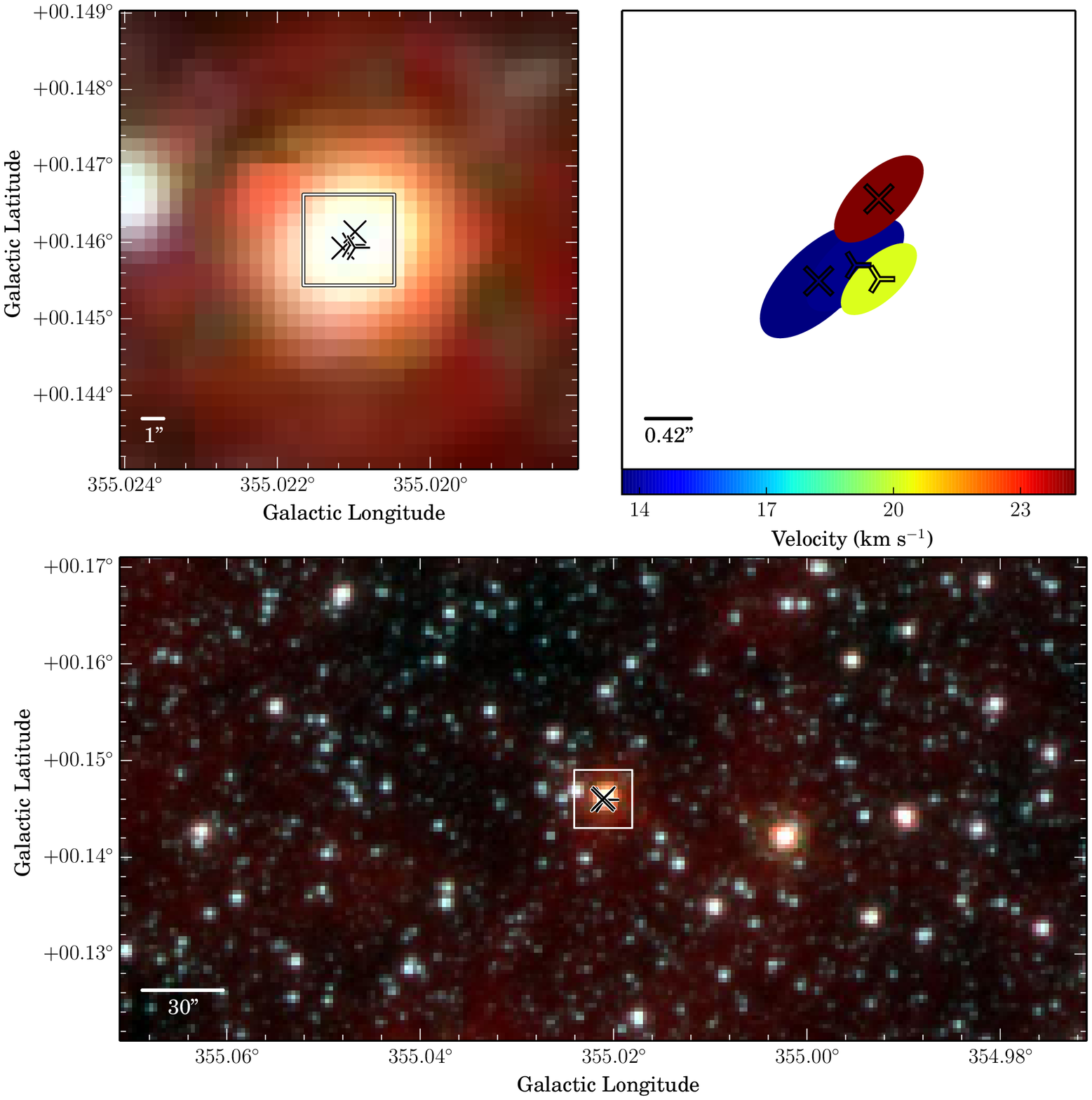}
\caption{G355.021$+$0.146 -- U. An example figure for one of the detected OH maser sites. Similar figures for each of the 356 detected OH maser sites are available online. In these figures, the upper panel shows the unbinned spectra of the OH maser transitions detected towards each site, with the radial velocity (with respect to the LSR) on the x-axis in units of \kms and flux density (derived from Stokes I) on the y-axis in units of Jy. The shaded regions of the spectrum represent the velocity range over which maser emission is detected. In the bottom, middle-left and middle-right panels of each of the figures, 1612 MHz maser spots are shown with 3-pointed stars, 1665 MHz maser spots with plus symbols, 1667 MHz maser spots with cross symbols and 1720 MHz maser spots with triangles. The bottom panel is a 6$\arcmin$ $\times$ 3$\arcmin$ three-color IR image, mostly from GLIMPSE (blue: 3.6\um; green: 4.5\um; red: 8.0\um) but supplemented by \textit{WISE} (blue: 3.4\um; green: 4.6\um; red: 12.2\um) for maser sites located beyond the GLIMPSE region. These images are centred on the presented maser site and are marked with all the maser spots detected within the range of the image. The white box shows the extent of the 21.6$\arcsec$ by 21.6$\arcsec$ region shown in the middle-left panel which shows only the spots from the single maser site. The middle-right panel shows the positions and relative error ellipses of all maser spots (derived by the \miriad task \textbf{imfit}) for the maser site, represented by both a colored symbol and a colored error ellipse (coded by velocity of the peak of each maser spot according to the color bar) which represents the relative positional uncertainty of the maser spot.}
\label{G355.021}
\end{figure*}

For each maser site, we present a figure showing the spectrum or spectra (depending on the number of transitions detected), the maser spots overlaid on the IR images and relative positional error ellipses for each maser spot (e.g., Figure \ref{G355.021}, which is in the same format as that in Figure 1 of \citealt{Qie2016a}). Note that the integrated flux density for each maser spot (listed in Table~\ref{maintable}) was derived from the area under the spectral line curve in the shaded channels of the spectrum which show the velocity range of the emission we have identified using the method outlined in Section \ref{observation}. In cases where significant emission appears in a spectrum but is not shaded, they have either been identified as noise spikes, or, more likely, are spectral features from nearby, but unrelated, strong masers (see descriptions for individual sources in Section \ref{individual}). Therefore, the shaded channels should be regarded as the velocity range for each maser spot we believe is real emission arising from the maser site. Note that six spectra (G003.951$+$0.262, G004.562$-$0.398, G004.680$+$1.498, G004.842$+$0.277, G004.962$-$0.017, G005.005$+$1.877) show zero values in some velocity ranges, which is caused by the flagged channels.

\subsection{Identification Criteria of Maser Sites}
\label{identification}

The SPLASH survey observed four ground-state OH transitions simultaneously with uniform sensitivity over a large area of the Galactic plane, thereby allowing us to derive the association properties of each of the OH maser transitions. In addition to comparing the associations between the different OH maser transitions we can also investigate the OH maser properties associated with different astrophysical objects. We note the possibility that the sensitivity limitations of the SPLASH survey could affect the numbers of OH masers identified at the locations of evolved stars and star formation differently, and therefore could have some effect on the following comparisons. 

As introduced in the pilot paper (\citealt{Qie2016a}), we can identify some OH maser sites based on their associations with class II methanol maser sites (\citealt{Cae2010}) since they are exclusively associated with high-mass star formation regions (e.g., \citealt{Bre2013}). The combination of a double-horned spectral profile at 1612 MHz and spatial coincidence with a bright star-like source in the GLIMPSE or WISE map indicates that the associated object is an evolved star. Since sources in this region around the Galactic center have, in general, more observational information available in the literature than those in the pilot region, we can use additional criteria. For instance, sources with identified variability of long cycle periods ($\ga 100$ days) in their optical/IR or OH maser emission, are evolved objects, such as Mira variables or OH/IR stars (e.g., \citealt{Gle2001}, \citealt{Vae1993}). Another useful indicator is the presence of SiO maser emission. So far, only seven confirmed regions of high-mass star formation are known to harbor this type of maser (see, e.g., \citealt{Ise2017} and references therein), so the SiO maser presence in a source is a good (although not definitive) indicator of an evolved star nature of the maser site. However, for the maser sites without obvious tracers described above, other methods are needed to identify their associations. Thus, we adopt several methods to classify each maser site into a wide variety of categories of the astrophysical associations, such as evolved stars including PNe, star formation, supernova remnants, or unknown objects. A detailed account of the steps taken to identify the astrophysical object associated with each of the maser sites is given in Section 4.1 in the pilot paper (\citealt{Qie2016a}). We employ the same steps in this portion of the Galaxy. However, in this region the Red MSX Source (RMS; \citealt{Lue2013}) catalogue could not be used because it does not cover the Galactic Centre region. The reason used for each of our astrophysical identifications is summarised for each source in Table \ref{maintable} in the second to last column. 

\subsection{Comments on Individual Sources of Interest}
\label{individual}

\noindent{\textit{G355.021$+$0.146}}. This maser site was also detected by \citet{Sea1997} with a single-peaked spectrum at 1612 MHz. \citet{Dee2004} detected three OH components towards this site. We detected two maser spots each at 1612 and 1667 MHz, as shown in Figure \ref{G355.021}. This site is probably associated with IRAS 17293-330, which fulfils the color criteria for a post-AGB star (\citealt{Dee2004}). In the GLIMPSE three-color image, the source seems to be associated with a star-like object. We did not find any clear identification for this source in the literature, thus it is designated as an unknown maser site.

\noindent{\textit{G355.156$-$0.597}}. The emission in the velocity range of $-$7\kms to $-$3\kms is from the nearby strong maser site G354.884$-$0.539 (not shown in this paper), whose flux density is about 100 Jy at $-$5\kms. 

\noindent{\textit{G355.292$-$0.240, G355.629$-$0.946, G355.873$+$0.086, G356.703$-$0.293, G357.405$-$1.206, G358.422$+$0.237, G358.623$-$1.730, G358.779$+$2.010, G358.831$-$0.271, G358.926$+$0.847, G358.936$-$1.078, G358.972$-$1.187, G359.054$-$0.114, G359.281$+$0.349, G359.404$+$0.860, G359.406$-$0.688, G000.007$-$0.817, G000.252$+$1.148, G000.647$+$1.890, G001.457$-$1.505, G001.648$+$1.177, G001.899$-$1.953, G002.031$-$1.635, G003.050$+$0.789, G003.415$-$0.309, G003.468$+$0.512, G003.650$+$0.787, G004.007$-$0.572, G004.703$+$1.552}}. These maser sites only exhibit one maser spot each at 1612 MHz. In the GLIMPSE three-color images, they are all associated with bright star-like objects, as shown in Figure 9 in \citet{Qie2016a}. Two of these sources (G359.406$-$0.688 and G003.468$+$0.512) are possibly associated with SiO maser emission (\citealt{Dee2000}), but given the SiO observations were conducted with the 45-m Nobeyama radio telescope, interferometric observations would be needed to confirm the associations. As described in the pilot paper, we also studied the properties of any IR point sources catalogued at the maser positions using GLIMPSE IR data. For those sources with IR point-source counterparts (7/29), the magnitude of the 4.5\um band is brighter than 7.8 for 5 sources (out of 7 sources, the remaining two sources do not have the 4.5\um magnitude), which suggests that these point sources are ``obscured'' AGB star candidates and are experiencing very high mass loss (\citealt{Roe2008}). However, although it is very possible that these maser sites originate from the circumstellar envelopes of evolved stars, we do not have any other independent confirmation of their nature. Thus we identify them as ``unknown maser sites''. 


\noindent{\textit{G355.344$+$0.147}}. This 1665 MHz OH maser is associated with a 6.7 GHz methanol maser (\citealt{Cae2010}), thus is identified to be associated with the star formation. This maser was detected with a clear Zeeman pattern of multiple features with a derived magnetic field of $-$3.4 to $-$5.4 mG by \citet{Fie2003} and $-$4.3 mG by \citet{Cae2013}. We did not detect the weak 1667 MHz OH maser reported by \citet{Cae2013} due to the lower sensitivity of our observations. This maser site is also associated with a 6035 MHz excited-state OH maser (\citealt{CV1995}; \citealt{Cas1997}) as well as a 22 GHz water maser (\citealt{Cae1983}; \citealt{Tie2016}).

\noindent{\textit{G355.944$-$0.041}}. This maser (associated with IRAS 17324-3221) is a re-detection of the maser in \citet{Cae1981}. It shows two maser spots at 1612 MHz and is associated with a star-like object in the GLIMPSE three-color image. No clear identification in the literature was found for this source, thus it was identified as an unknown site.

\noindent{\textit{G356.457$-$0.386}}. This source was identified as an evolved star site based on \citet{Bee2003}. It was also detected by \citet{Sea1997} with an irregular spectrum (defined as neither single- or double-horned) at 1612 MHz. We detected four maser spots at 1612 MHz, also with an irregular spectrum. 

\noindent{\textit{G356.568$+$0.318}}. This maser source fulfils the color criteria for AGB stars in \citeauthor{Mee2002} (\citeyear{Mee2002}), (\citeyear{Mee2004}), (\citeyear{Mee2005}). It is also variable in the IR flux densities (\citealt{Mee2002}), which further supports its AGB nature. 

\begin{figure*}
\includegraphics[width=0.9\textwidth]{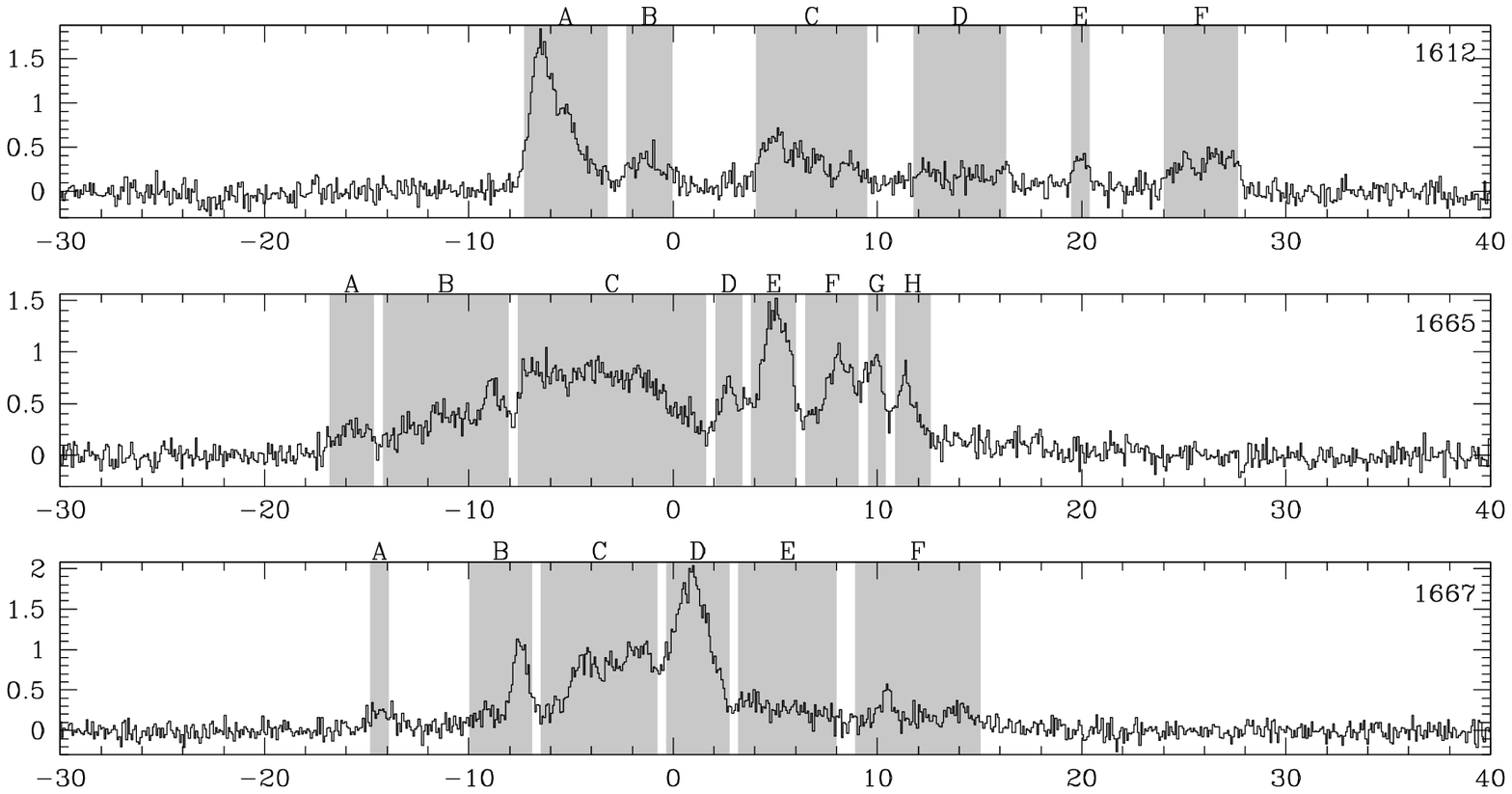}
\includegraphics[width=0.9\textwidth]{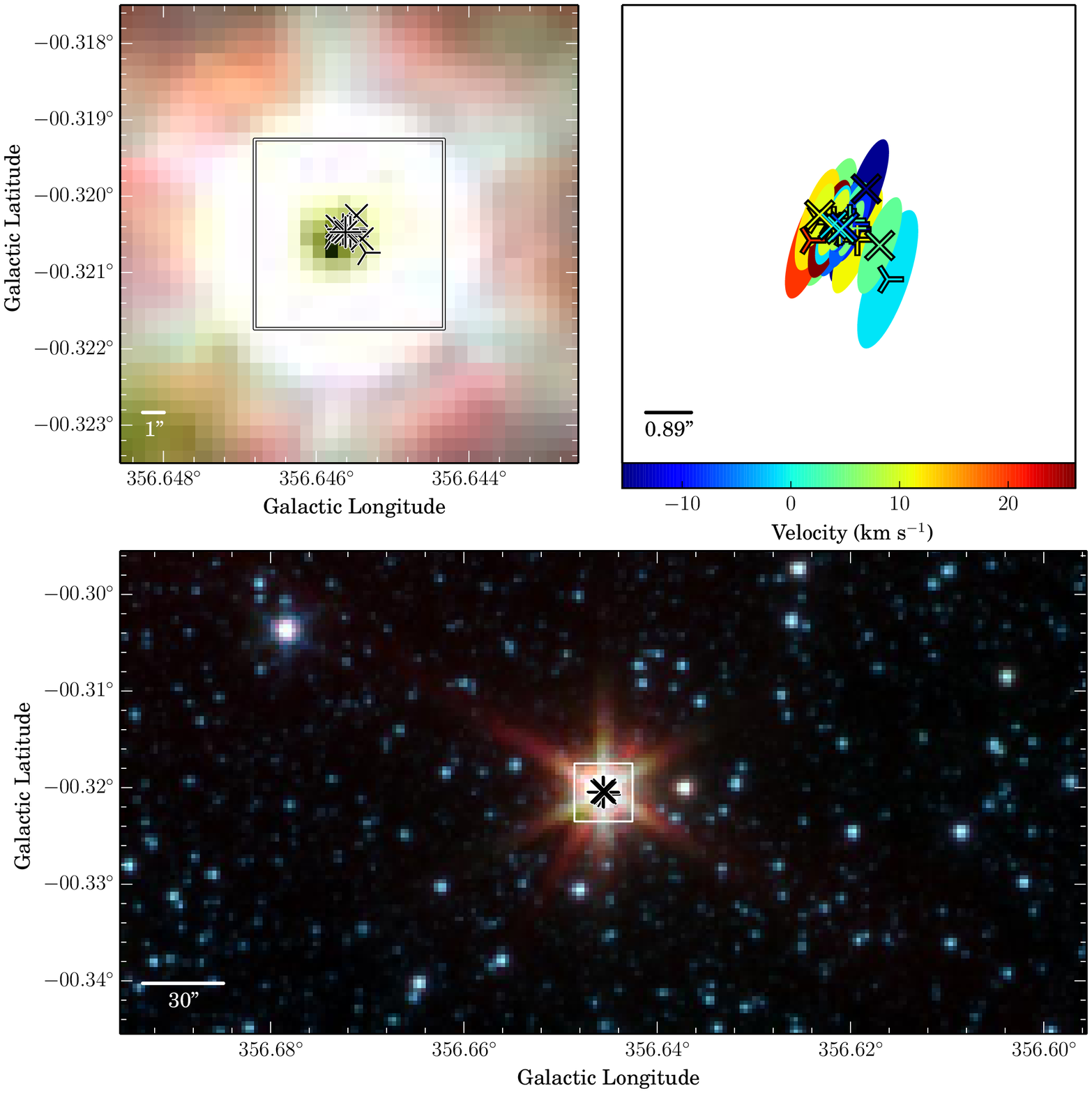}
\caption{G356.646$-$0.321 -- ES}
\label{G356.646}
\end{figure*}

\noindent{\textit{G356.646$-$0.321}}. This evolved star maser site (IRC-30308, IRAS 17354-3155; Figure \ref{G356.646}) exhibits 20 maser spots at 1612, 1665 and 1667 MHz and is the richest maser site in the Galactic Center region. The 1665 and 1667 MHz transitions are at the same velocity range (from $-$18\kms to $+$15\kms), whereas the 1612 MHz transition also shows maser emission in the velocity range of $+$14\kms to $+$28\kms. Given its rich spectra at 1612, 1665 and 1667 MHz (clearly departing from the double-horned profile), this evolved star is probably a post-AGB star. A SiO maser has been detected towards this source (\citealt{Che1996}, \citealt{Lee1978}). The GLIMPSE three-color image is saturated at the center.

\noindent{\textit{G356.646$-$0.153, G357.179$-$0.521, G357.749$+$0.320, G357.908$+$0.234, G357.988$-$0.988, G358.426$-$0.175, G358.726$-$0.268, G359.033$+$1.938, G359.140$+$1.137, G359.201$+$0.285, G359.233$-$1.876, G359.284$+$0.247, G359.360$+$0.084, G359.512$-$0.659, G359.732$+$1.260, G359.899$+$0.222, G000.260$+$1.026, G000.452$-$1.216, G000.484$-$0.167, G000.494$-$0.211, G000.699$-$1.191, G000.762$+$0.768, G000.814$+$0.179, G001.212$+$1.257, G001.227$+$2.005, G001.233$+$1.273, G001.484$-$0.061, G001.620$-$1.560, G001.794$+$2.078, G001.817$+$1.988, G002.088$-$1.047, G002.232$+$0.016, G002.286$-$1.801, G003.117$+$0.682, G003.203$+$0.024 , G003.942$-$0.007, G004.680$+$1.498}}. These are evolved star maser sites, which exhibit the typical double-horned profiles at 1612 MHz. Several sites also show double-horned profiles at 1667 MHz (G359.233$-$1.876, G000.452$-$1.216, G001.212$+$1.257, G001.484$-$0.061 and G003.203$+$0.024). In GLIMPSE three-color images, these stars are very red (bright at 8\um).

\noindent{\textit{G356.840$-$0.045}}. This maser site has been detected by \citet{Tae2009} with a double-horned profile at 1612 MHz (as source 2MASS J17380406-3138387). This maser is associated with a bright star-like GLIMPSE source. Thus, combining Tafoya's spectrum and the IR image, we include this site in the evolved star category. Note that we only detected the blue-shifted 1612 MHz maser component at the velocity of about $-$120\kms.

\noindent{\textit{G357.208$+$1.747}}. This maser site only has one maser spot at 1612 MHz. In the absence of confirmation of its nature in the literature, we identify it as an unknown site. In the WISE three-color image, it is associated with a bright star-like object. 

\begin{figure*}
\includegraphics[width=0.9\textwidth]{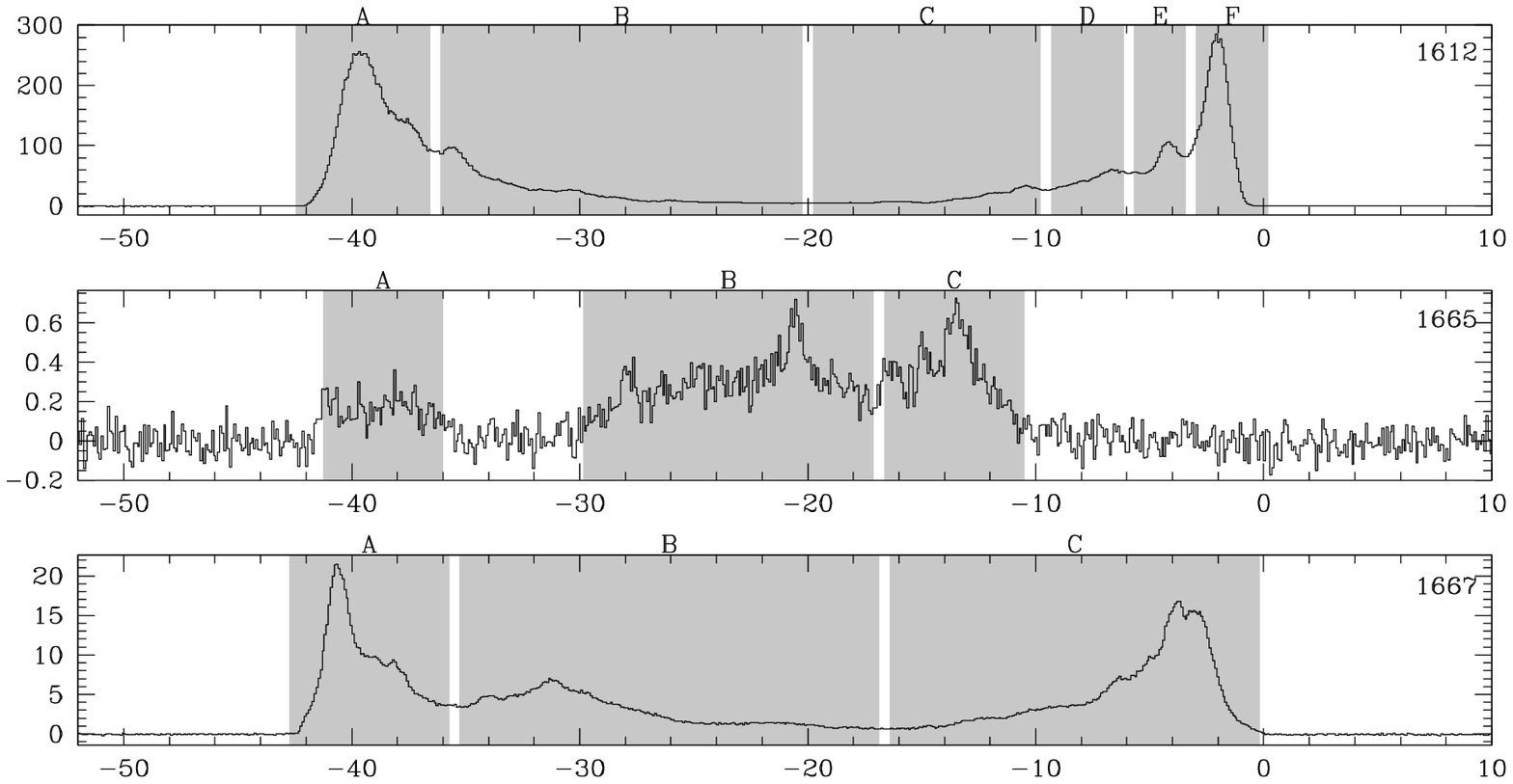}
\includegraphics[width=0.9\textwidth]{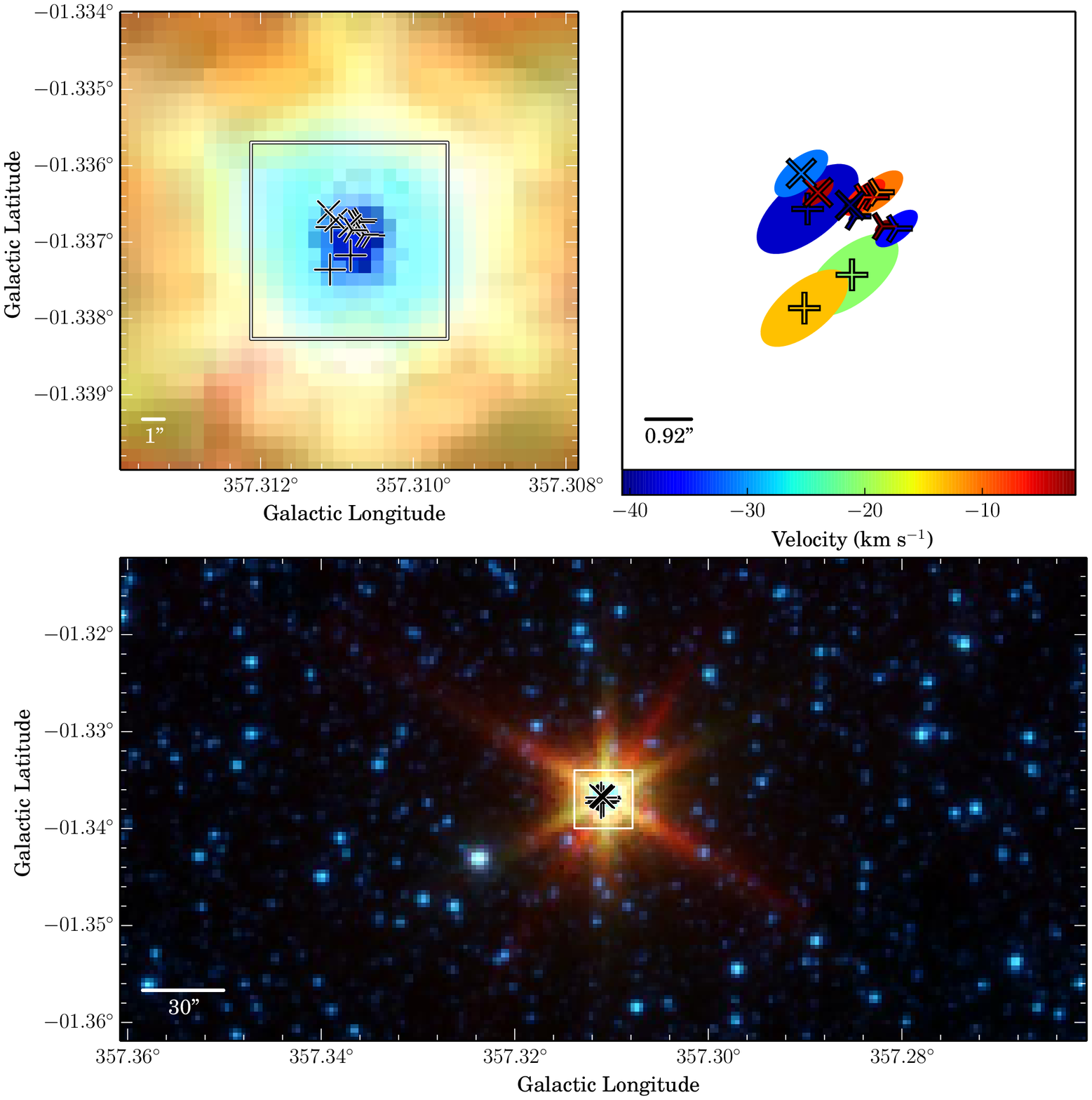}
\caption{G357.311$-$1.337 -- ES}
\label{G357.311}
\end{figure*}

\noindent{\textit{G357.311$-$1.337}}.  This maser site is the strongest maser site in the SPLASH survey region (shown in Figure \ref{G357.311}). The peak flux density of the 1612 MHz transition is about 285 Jy. The 1612 and 1667 MHz spectra are similar. This site is associated with an evolved star and the GLIMPSE three-color image is saturated at the center.

\noindent{\textit{G357.473$+$0.367}}. This source was identified as an evolved star site by \citet{Sea1997}. It has five maser spots in the 1612 MHz spectrum. In the GLIMPSE three-color image, its associated star-like object is located within diffuse IR background.

\noindent{\textit{G357.637$+$0.293}}. The unassociated emission in the 1612 MHz spectrum is from a nearby source, G357.473$-$0.367.

\noindent{\textit{G357.653$-$0.056}}. This source only has one maser spot at 1720 MHz. The linewidth of the maser spot is about 1\kms. This maser has been detected by previous surveys, e.g., \citet{Fre1996} and \citet{Yue1999}, which searched 1720 MHz OH maser emission around SNRs. \citet{Yue1999} argued that this 1720 MHz OH maser is associated with the Tornado nebula. However, there is still a debate on the nature of the Tornado nebula. Thus we identify this maser site as having unknown origin.

\noindent{\textit{G357.730$+$1.681}}. This maser site is associated with a Mira type variable star V2488 Oph (\citealt{Sae2003}). The 1612 MHz OH maser spectrum has three maser spots distributed in the velocity range of $+$64\kms to $+$74\kms. The 1665 MHz OH maser spectrum only exhibits one maser spot at $+$70\kms.

\noindent{\textit{G358.083$+$0.137}}. The unassociated emission at $-$18\kms is from the nearby source G358.162$+$0.490. 
 
\noindent{\textit{G358.235$+$0.115}}. This maser site (associated with IRAS 17376-3021) is an OH/IR star site (\citealt{Wae2014}), and was detected by \citet{Sea1997} with an irregular spectrum at 1612 MHz. In our observations, the 1612 MHz spectrum is also irregular with seven maser spots distributed in the velocity range of $-$32\kms to $+$18\kms. Four maser spots are detected in the 1665 MHz transition and five maser spots are in the 1667 MHz transition. This site also showed strong water maser emission (over 70 Jy) over a velocity range of about 40\kms (\citealt{Cae1983}; \citealt{FC1989}). \citet{Wae2014} also re-detected this strong water maser emission in their observations. Considering its unusual OH and water maser characteristics, this star is probably a post-AGB star, with the masers tracing non-spherical mass-loss. Moreover, \citet{Kie2013} found a SiO maser towards this site. In the GLIMPSE three-color image, the center pixels are saturated. Note that the absorption features in the 1612 MHz spectrum are sidelobe contamination from the nearby source G358.162$+$0.490. 

\noindent{\textit{G358.291$+$0.081}}. This OH maser is a new detection, which has been identified as an evolved star by HOPS (\citealt{Wae2014}). It shows one maser spot at 1612 MHz, two maser spots at 1665 MHz and three maser spots at 1667 MHz. The center of the GLIMPSE three-color image is saturated.

\noindent{\textit{G358.359$+$0.088}}. In the SIMBAD, a star (IRC-30314, \citealt{HB1975}; also IRAS 17380-3015) is located at this position. \citet{HB1975} did not verify whether this star is an evolved star. The SiO maser emission (\citealt{Hae1990}) and the water maser emission (\citealt{Tae1993}) have also been detected towards this source. However, the SiO maser observations were made with a single-dish telescope so interferometric observations would be needed to verify this association. In addition, the 1612 MHz OH maser spectrum does not show the typical double-horned spectral profile, thus we identified this source as an unknown maser site. Two 1665 MHz maser spots and three 1667 MHz maser spots are also detected. The unassociated emission at $+$0\kms in the 1612 MHz spectrum is from the nearby source G358.235$+$0.115. The center of the GLIMPSE three-color image is saturated.

\noindent{\textit{G358.394$-$0.284}}. In SIMBAD\footnote{http://simbad.u-strasbg.fr}, we find that a star (IRAS 17396-3025) is close to this maser site position. However, this maser site only shows two maser spots at 1665 MHz. Thus, we classify this source as an unknown site. Note that in the GLIMPSE three-color image, the associated object is very red (bright at 8\um). 

\noindent{\textit{G358.656$-$1.710}}. This maser site is a new detection. It is very close to G358.649$-$1.701 (with an angular separation about 41\arcsec). This site shows one maser spot at 1665 MHz and one maser spot at 1667 MHz. We searched the literature, compared the coordinates and identified this maser site to be associated with a variable star (Terz V 2118; \citealt{TO1988}). In the GLIMPSE three-color image, this site looks saturated in the center.

\noindent{\textit{G358.831$-$0.175}}. This maser site is associated with an AGB star (\citealt{Sce2003}). It only exhibits one maser spot in the 1612 MHz transition. 

\noindent{\textit{G358.936$-$0.485, G359.145$-$0.356}}. These two 1720 MHz OH maser sites have been detected by \citet{Yue1995} as source C2 and A, respectively. According to these authors and \citet{WY2002}, these two 1720 MHz OH masers are associated with the SNR G359.1$-$0.5. Thus, we classified them into the SNR category. Notably, the peak velocities of these two masers are $-$6.3\kms and $-$5.1\kms in our observations, and $-$5.57\kms and $-$4.47\kms in \citet{Yue1995}. In the GLIMPSE three-color images, they are located in the IR extended emission background.

\noindent{\textit{G358.983$-$0.652}}. This 1720 MHz OH maser has a peak flux density of about 2.2 Jy at $-$5.9\kms. This detection seems to correspond to the southernmost maser spot in Fig 3A of \citet{WY2002}, which lies at the edge of the SNR G359.1$-$0.5. We believe it was the maser listed as source D in \citet{Yue1995}, although there must be a typographical error in their Table 1, and its B1950 declination should be $-$30:07:24.11 rather than $-$30:04:24.11. Given the information above, we classified this source into the SNR category. In the GLIMPSE three-color map, this site is located in the diffuse emission background.

\noindent{\textit{G358.987$+$1.131, G359.487$-$1.067, G004.372$+$.058}}. Although these maser sites show a double-horned profile in the 1612 MHz spectra, they are not associated with any star-like objects in the GLIMPSE three-color images. We searched the literature, but found no objects are located at these positions. Therefore, we classified these sources as unknown maser sites. 

\noindent{\textit{G359.150$-$0.043}}. This 1612 MHz OH maser is associated with an OH/IR star and was detected with a double-peaked profile by \citet{Sea1997}. In our observations, we detected three maser spots distributed in the velocity range of $+$34\kms to $+$43\kms.

\noindent{\textit{G359.230$-$1.309}. This 1612 MHz OH maser is a new detection with a peak flux density of 0.32 Jy at $+$70.6\kms. We did not find any associated object for this source in the literature. In the GLIMPSE three-color image, it is located in the diffused background. Thus, we identified this maser site as an unknown site.

\noindent{\textit{G359.260$+$0.164}}. In the literature, this maser (associated with IRAS 17400-2927) originates from a variable star (\citealt{Sce2000}), thus we identify this source as an evolved star site. This source has also been detected with a 22 GHz water maser in the velocity range of $+$54.8\kms to $+$60.1\kms with a peak flux density of 1.20 Jy (\citealt{Tae1993}). In our observations, we detected two maser spots in the velocity range of $+$50.3\kms to $+$57.3\kms in the 1612 MHz transition. Moreover, this site has also been detected with a 43 GHz SiO maser peaked at $+$75.5\kms (with the Nobeyama telescope; \citealt{Fue2006}). The velocity of the SiO maser (tracing the stellar velocity) does not seem to be consistent with those of OH masers and water masers (the velocity difference is about 20\kms), thus we are not certain whether the SiO maser is associated with our OH masers. 

\noindent{\textit{G359.380$-$1.201}}. This maser site is an evolved star site (\citealt{Sea1997}). The 1612 MHz spectrum is double-horned, but deviates from the classical profile shape, in that it has multiple components corresponding to six maser spots. The 1665 MHz profile is also double-horned, with a suggestion of similarly complex structure. Notably, the 1665 MHz peaks (at $-$228 and $-$209\kms) are at less extreme velocities than the 1612 MHz peaks (the brightest of which are at $-$235\kms and $-$203\kms), and have no obvious counterparts in the 1612 spectrum. This is expected, since the 1665 MHz maser tends to arise from an innermost region compared to the 1612 MHz OH maser (e.g., \citealt{DT1991}; \citealt{Coh1993}; \citealt{Dee2004}). In the GLIMPSE three-color image, this site is very bright at 8\um.

\noindent{\textit{G359.429$+$0.035}}. This 1612 MHz OH maser has been detected by \citet{Lie1992} and \citet{Sea1997}. In the spectrum of \citet{Lie1992}, three peaks in the velocity range of $-$16\kms to $+$20\kms were detected and the middle broad component might correspond to what we have detected at $+$2.1\kms. \citet{Sea1997} detected a double-peaked profile at 1612 MHz (peaked at $-$10.3\kms and $+$1.3\kms) and classified this source as an evolved star site. Thus, we include this source into the evolved star category. Note that the only maser spot we detect has a broad linewidth, which is about 11\kms. In the GLIMPSE three-color image, this site is located in the diffuse background.

\noindent{\textit{G359.443$-$0.840}}. In \citet{Zie1989} and \citet{Rae1990}, this maser site (associated with IRAS 17443-2949) was associated with a PN. However, \citet{Goe2008} did not find any continuum emission associated with this source, and pointed out that the position of the continuum source reported by \citet{Rae1990} is not compatible with the position of the OH masers. \citet{Use2012} also extensively discussed this source and did not include it into the OH-maser-emitting PN (OHPN) category. No optical counterpart was detected by \citet{Sue2006}. In \citet{Rae2012}, the source is marginally extended in the near-IR, but in our GLIMPSE three-color image, this source shows the extended structure and is very bright at 8\um. Moreover, this source fulfils the color criteria of a post-AGB star in \citet{Sue2006}. For the reasons above, we include this source into the evolved star category (a post-AGB star). In the spectrum of \citet{Zie1989}, only one 1612 MHz maser spot with a peak at $-$15\kms was detected. In our observations, we detected a double-horned profile in the 1612 MHz transition. The velocity range of the 1612 MHz OH maser is between $-$18.6\kms and $+$11.9\kms. Three 1665 MHz OH maser spots in the velocity range of $-$18.7\kms to $-$10.4\kms and one 1667 MHz maser spot peaked at $-$17.7\kms were also detected. \citet{Goe2008} detected 1612 and 1665 MHz OH masers as well as a water maser towards this source with the VLA.

\noindent{\textit{G359.445$-$0.267}}. This maser site has been detected with a double-horned profile at 1612 MHz (peaked at $-$125.1\kms and $-$90.0\kms) and was identified as an OH/IR star by \citet{Lie1992}. Thus, we classify it as an evolved star. In our observations, only the maser spot in the red-shifted velocity range (peaked at $-$89.4\kms) is detected. In the GLIMPSE three-color image, this site is associated with a bright star-like object.

\noindent{\textit{G359.543$+$1.776}}. This maser site is an evolved star site (\citealt{Sea1997}). The 1665 MHz spectrum peaks at similar velocities to the 1612 MHz spectrum. The 1612 MHz and 1665 MHz spectra deviate from the typical double-horned profiles. In the GLIMPSE three-color image, this source is very red (bright at 8\um).  

\noindent{\textit{G359.567$+$1.147}}. This maser site shows a double-horned profile at 1612 MHz. In the GLIMPSE three-color image, there is a faint and red object, which could be its counterpart. A continuum source at 330 MHz is located at 6\arcsec\ away from this OH maser with an astrometric error of about 2\arcsec\ (\citealt{Lae2000}; \citealt{Noe2004}). Thus, it is interesting to check the continuum emission in our data. Given the information above, we include this source into the unknown category.

\noindent{\textit{G359.581$-$0.240}}. The 1612 MHz OH maser at this site is a re-detection of the 1612 MHz OH maser in \citet{Sea1997} and has been identified as an OH/IR star site by \citet{Hae1983}. In our observations, the maser site shows four maser spots at 1612 MHz, two maser spots at 1665 MHz and one maser spot at 1667 MHz. In the GLIMPSE three-color image, this maser site is located in the diffused background.

\noindent{\textit{G359.858$+$1.005}}. This 1612 MHz OH maser was detected with a double-peaked profile by \citet{Sea1997}, and was identified as an evolved star site. However, in our observations, only the red-shifted maser spot peaked at $-$33.6\kms was detected.

\noindent{\textit{G359.932$-$0.063}}. This 1612 MHz OH maser was identified as a star formation OH maser site, since it is associated with a young stellar object (YSO; \citealt{Yue2009}). It shows a single peak of 1.5 Jy at $-$113.2\kms and has also been detected by \citet{Lie1992}. In the GLIMPSE three-color image, this maser site is associated with an extended green object (EGO; \citealt{Cye2008}).

\noindent{\textit{G359.939$-$0.052}}. This 1612 MHz OH maser site was identified as an evolved star site by \citet{Lie1992} and \citet{Sje1998}. This maser site was only identified when the ATCA images were added as introduced in Section \ref{observation}. We detected one 1612 MHz maser spot at the velocity of $+$69.9\kms with a linewidth of about 1\kms. The star associated with this maser is a variable star -- V4513 Sgr, with a periodicity of about 400 days both in the IR band (\citealt{Gle2001}) and the OH maser emission (\citealt{Vae1993}). Thus, this star is an OH/IR star. \citet{Yue2015} also detected a water maser with two peaks at $+$38.84\kms and $+$67.55\kms, which is consistent with the velocities of our OH maser. In the GLIMPSE three-color image, this site is associated with a faint IR star. 

\noindent{\textit{G359.940$-$0.067}}. This 1720 MHz OH maser has been detected by \citet{Yue1996} without any classification. \citet{Yue1999} and \citet{Yue2001} showed this maser is associated with a SNR shell (Sgr A East). Thus, we assigned this maser site to the SNR category. \citet{Yue1996} inferred a magnetic field strength of about $+$3.7 mG in the line-of-sight direction from the Zeeman effect in the 1720 MHz maser. In the GLIMPSE three-color image, this site is located in the diffuse background. 

\noindent{\textit{G359.951$-$0.036}}. This maser site was identified as an evolved star site by \citet{Lie1992}. \citet{Sea1997} detected a singe-peaked spectrum towards this source. We re-detected this maser component at about $+$96\kms. With the VLA, \citet{Sje2002} also detected a 43 GHz SiO maser at a velocity of about $+$84.9\kms (tracing the stellar velocity) towards this site, which is 2.25\arcsec\ away from our position. Thus, it is likely that the SiO maser and the OH maser originate from the same source. In the GLIMPSE three-color image, this maser site is located in the diffuse background.

\noindent{\textit{G359.953$-$0.036}}. This 1720 MHz OH maser has also been detected by \citet{Yue1996}, who suggested that this 1720 MHz OH maser is physically associated with Sgr A East (a non-thermal radio source). \citet{Yue1996} inferred the magnetic field with this 1720 MHz OH maser, resulting in about $+$2.0 mG in the line-of-sight direction. Based on previous studies, we classified this source as an unknown site. In the GLIMPSE three-color image, this source is located in the extended emission background.

\noindent{\textit{G359.956$-$0.041}}. This 1720 MHz OH maser site has also been detected by \citet{Yue1996} and \citet{Yue1999}, who considered that this maser site is physically associated with Sgr A West (a thermal source). They obtained a magnetic field strength of about $-$4.0 mG in the line-of-sight direction. They did not identify this maser site, thus we assigned this maser as an unknown maser site. In the GLIMPSE three-color image, this site is located in the extended emission background.

\noindent{\textit{G359.966$-$1.144}}. This 1612 MHz OH maser (associated with IRAS 17468-2932) is a new detection. According to \citet{Mae2005a} and \citet{Soe2013}, this site is associated with a variable star of the Mira type with a periodicity of about 600 days, thus we classified this site into the evolved star category. \citet{Deg2004} detected a SiO maser at $+$52.9\kms towards this source with the Nobeyama telescope. In \citet{JV2004}, this source is a PN candidate, Jast2 6, which is identified based on the extended H$\alpha$ emission. \citet{Mie2009} considered this source as a PN mimic (possibly a symbiotic star). This maser only shows one maser spot at a peak velocity of $+$41\kms with a flux density of 1.4 Jy. The linewidth of this maser spot is about 5\kms. 

\noindent{\textit{G000.074$+$0.145}}. This 1612 MHz OH maser site was identified as an evolved star site by \citet{Lie1992} and \citet{Sje1998}. This maser site was only identified once the ATCA images were added as introduced in Section \ref{observation}. We only detected one maser spot towards this source with a linewidth of about 0.4\kms. This maser site may be associated with a long period variable V4489 Sgr (about 8\arcsec\ away), with a periodicity of 645 days in the IR band (\citealt{Joe1994}) and 639 days in the OH maser emission (\citealt{Vae1993}). In the GLIMPSE three-color image, this site is associated with a faint IR star. 

\noindent{\textit{G000.170$+$0.534}}. This 1667 MHz OH maser is a new detection. In the literature, we find an IR-red source (IRAS 17407-2829) is associated with this maser site. We use double-horned profiles at 1612 MHz as a criterion to classify a source as an evolved star (combined with a star-like object in the IR image). Despite the double-horned profile at 1667 MHz, since this site is not detected at 1612 MHz, we classify this source as an unknown site.

\noindent{\textit{G000.178$-$0.055}}. This maser site shows one maser spot at 1612 MHz and one maser spot at 1667 MHz. These two maser spots are in a similar velocity range. \citet{Lie1992} classified this source as an OH/IR star. Thus, we assigned this maser into the evolved star category. \citet{Vae1993} obtained an OH maser periodicity of 551 days. \citet{Sje2002} found this source might be associated with a 22 GHz water maser. In the GLIMPSE three-color image, this source is associated with a star-like object.

\noindent{\textit{G000.189$+$0.053}}. This maser site is associated with a variable star V4524 Sgr, which is also an OH/IR star (\citealt{Lie1992}). The periodicity of OH masers is about 843 days (\citealt{Vae1993}) and the periodicity of the IR flux densities is about 885 days (\citealt{Woe1998}). Only one maser spot at 1612 MHz was detected by our observations. In the GLIMPSE three-color image, this site is associated with a star-like object.

\noindent{\textit{G000.207$+$1.413}}. This maser site (associated with IRAS 17375-2759) was a PN candidate (\citealt{Use2012}). It has OH maser emission and radio continuum emission, but its nature as a PN has not been confirmed spectroscopically (\citealt{Use2012}). Thus, we included this source in the evolved star category (\citealt{Sea1997}). \citet{Sea1997} detected two maser spots at 1612 MHz towards this site. In our observations, we only detected the blue-shifted component in the velocity range of $+$19.7\kms to $+$25.1\kms. We also detected a weak maser spot at 1667 MHz. In the GLIMPSE three-color image, this evolved star is very red (bright at 8\um).

\noindent{\textit{G000.319$-$0.040}}. This maser site is associated with an evolved star (\citealt{Sea1997}). The two maser peaks reported by \citet{Sea1997} were also detected by our observations. In fact, we detected four maser spots in total at 1612 MHz. In the GLIMPSE three-color image, the associated star-like object is very red (bright at 8\um).

\noindent{\textit{G000.333$-$0.180}}. This maser site has been detected by \citet{Sea1997} with a single-peaked spectrum. In our observations, we detected the typical double-horned profile with four maser spots in the 1612 MHz transition. We searched the literature and assigned this source to the evolved star category based on \citet{Boe1978}. This maser site is located in the diffuse background in the GLIMPSE three-color image. 

\noindent{\textit{G000.344$+$1.566}}. This maser site is a well-known OHPN (JaSt 23; \citealt{Vae2001}). The velocity ranges of the 1612 MHz and 1665 MHz OH masers are very broad -- about 15\kms and 11\kms, respectively, as shown in Figure \ref{G000.344}. \citet{Sea1997} detected a single-peaked spectrum at 1612 MHz, the peak velocity of which is $+$115.2\kms. We also detected this peak in our observations. \citet{Goe2016} detected a possible Zeeman Pair at 1665 MHz with a derived magnetic field of B = 0.8-3 mG. In the GLIMPSE three-color image, this PN appears very bright at 8\um, thus very red.

\begin{figure*}
\includegraphics[width=0.9\textwidth]{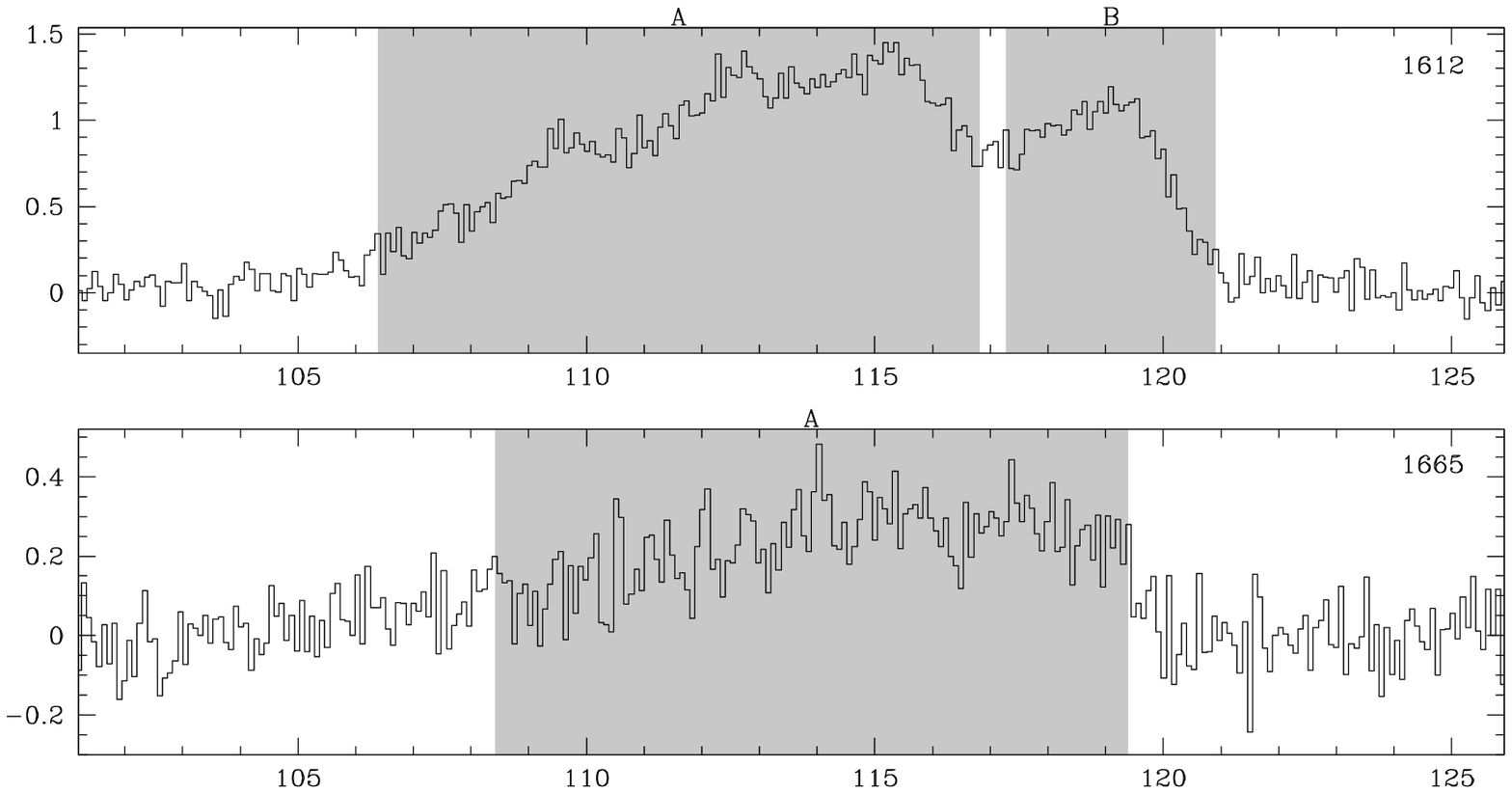}
\includegraphics[width=0.9\textwidth]{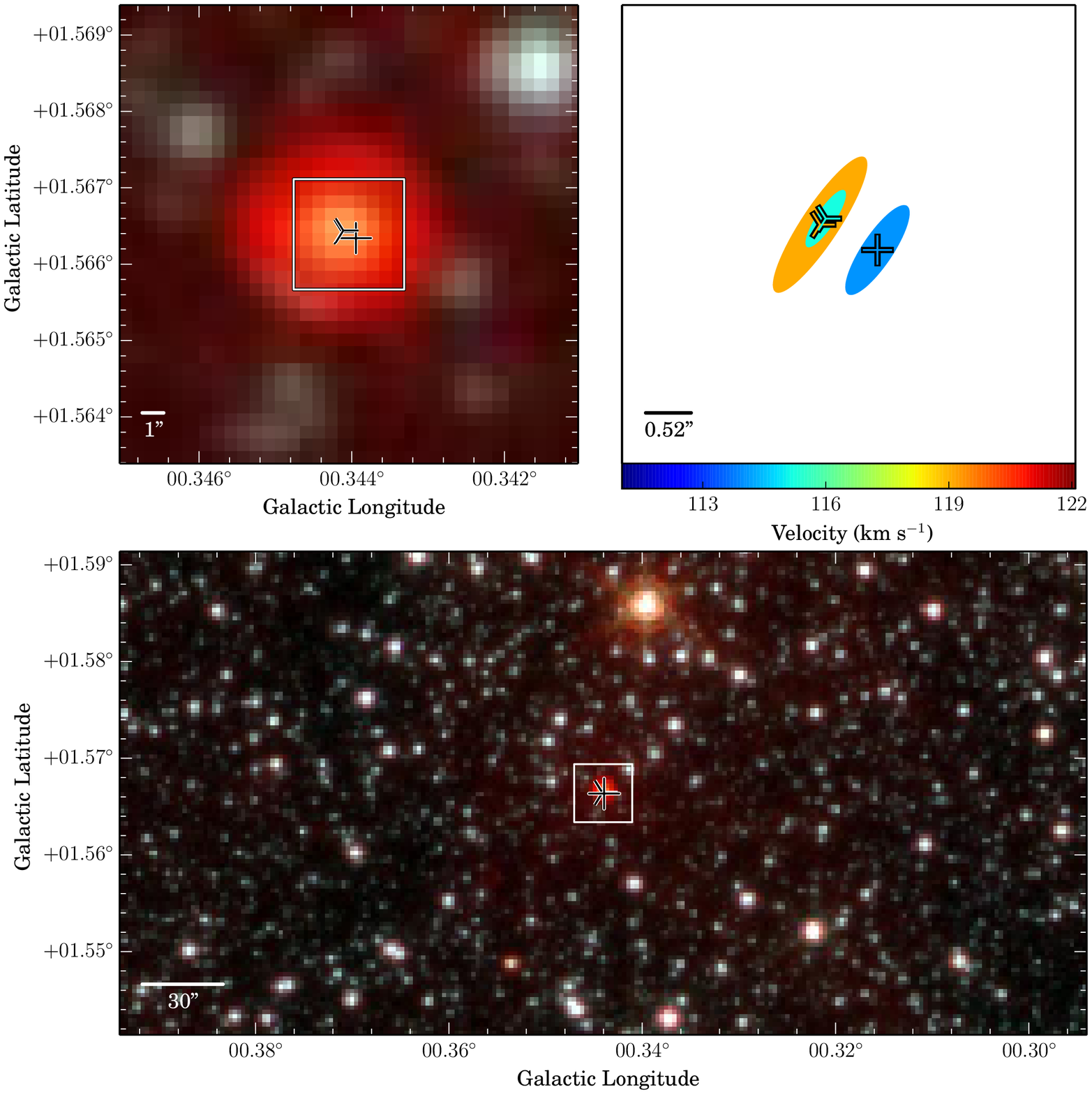}
\caption{G000.344$+$1.566 -- PN}
\label{G000.344}
\end{figure*}

\noindent{\textit{G000.453$+$0.026}}. This maser site is associated with an OH/IR star with a double-horned profile at 1612 MHz (\citealt{Lie1992}). In our observations, only the red-shifted maser spot at 1612 MHz was detected. The linewidth of the maser spot is about 1\kms. In the GLIMPSE three-color image, this maser site is associated with a very red star-like object (bright at 8\um).

\noindent{\textit{G000.517$+$0.050}}. This maser was identified as an evolved star site by \citet{Sea1997}. \citet{Sea1997} detected a double-peaked spectrum at 1612 MHz. We only detected the red-shifted component at $+$184.6\kms. 
In the GLIMPSE three-color image, this maser site is associated with a very red star (bright at 8\um).

\noindent{\textit{G000.647$+$1.890}}. This maser site (associated with IRAS 17367-2722) has also been detected by \citet{Sea1997} with a single-peaked spectrum at $+$52.5\kms. We re-detected this 1612 MHz OH maser component. After searching the literature, we found this source was detected with two peaks at 1612 MHz both by \citet{Tee1991} and \citet{Dae1993} (peaked at $+$37.1\kms and $+$52.5\kms). However, both the observations of \citet{Tee1991} and \citet{Dae1993} were conducted with single-dish telescopes. Thus, we are not certain whether these two peaks originate from the same maser site. Given the information above, we classify this source as an unknown maser site. In the GLIMPSE three-color image, this source is associated with a very red star-like object (bright at 8\um).

\noindent{\textit{G000.657$-$0.040}}. This 1665 MHz OH maser was detected by \citet{CH1983a} (source 0.66-0.04) and it is also associated with a 6.7 GHz methanol maser (\citealt{Cae2010}). In the GLIMPSE three-color image, this maser site is close to an EGO. The unassociated emission in the 1665 MHz spectrum is from the nearby strong star formation maser source G000.667$-$0.035.

\noindent{\textit{G000.658$-$0.042}}. This maser site is the strongest star formation OH maser site in the Galactic Center region. The associated star formation site is named Sgr B2S. The peak flux density of the 1665 MHz OH maser can be as strong as 154 Jy. It was also detected by \citet{Cas1998}. In the GLIMPSE three-color image, this maser site is associated with a bright EGO.

\noindent{\textit{G000.666$-$0.035}}. This 1720 MHz OH maser site is associated with a 6.7 GHz methanol maser from the MMB survey (\citealt{Cae2010}), thus was identified as a star formation site. This site is the well-studied star formation site Sgr B2M. \citet{Are2000} also detected this 1720 MHz OH maser with the Very Large Array (VLA). In \citet{Cas2004}, there were obvious absorption features at about $+$64\kms in this 1720 OH maser spectrum and we also detected these absorption features in our observations. In the GLIMPSE three-color image, this site is close to several bright EGOs.

\noindent{\textit{G000.666$-$0.029}}. This maser site contains two maser spots at 1612 MHz, which are also associated with a 6.7 GHz methanol maser. This 1612 MHz OH maser is a new detection and shows absorption features at a velocity lower than $+$70\kms. In the GLIMPSE three-color image, this site is located in the diffuse background. 

\noindent{\textit{G000.667$-$0.035}}. This maser site belongs to the star formation region Sgr B2M. These two 1612 MHz OH maser spots have been detected by \citet{Sea1997}. The 1665 and 1667 MHz OH masers were detected by \citet{Cas1998}. This maser site also shows 6.7 GHz methanol masers. In the GLIMPSE three-color image, these maser spots are associated with EGOs. The unassociated emission in the 1665 and 1667 MHz spectra is from the nearby strong star formation maser site G000.658$-$0.042.

\noindent{\textit{G000.667$-$0.036}}. This 1720 MHz OH maser also belongs to the star formation region Sgr B2M. In the literature, this maser site is associated with Sgr B2M \hii region (\citealt{Bue1990}). This maser is a new detection and shows absorption features at velocities higher than $+$59\kms. In the GLIMPSE three-color image, this site is associated with an EGO, which is located at a lower Galactic latitude than the EGO associated with G000.667$-$0.035 (see Figure \ref{G000.667}). 

\begin{figure*}
\includegraphics[width=0.9\textwidth]{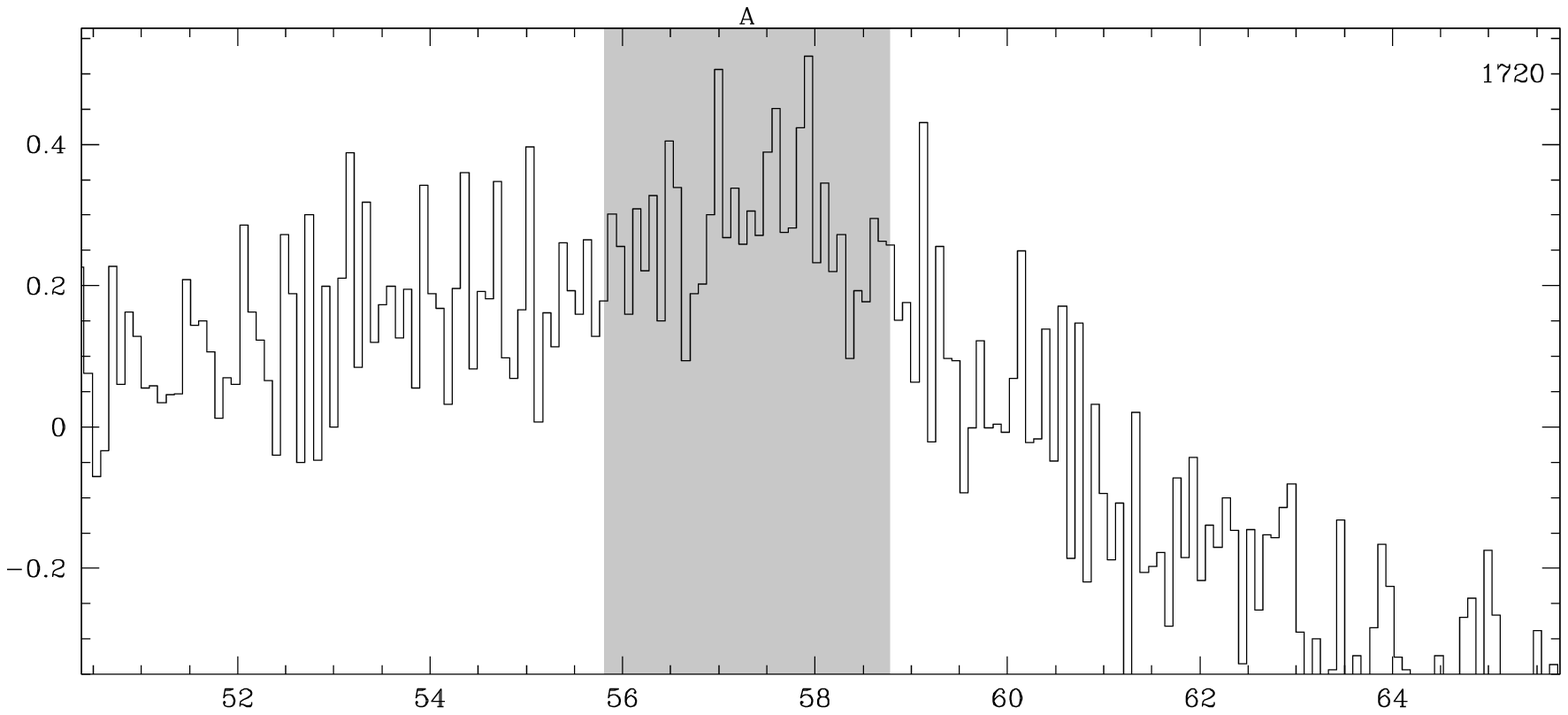}
\includegraphics[width=0.9\textwidth]{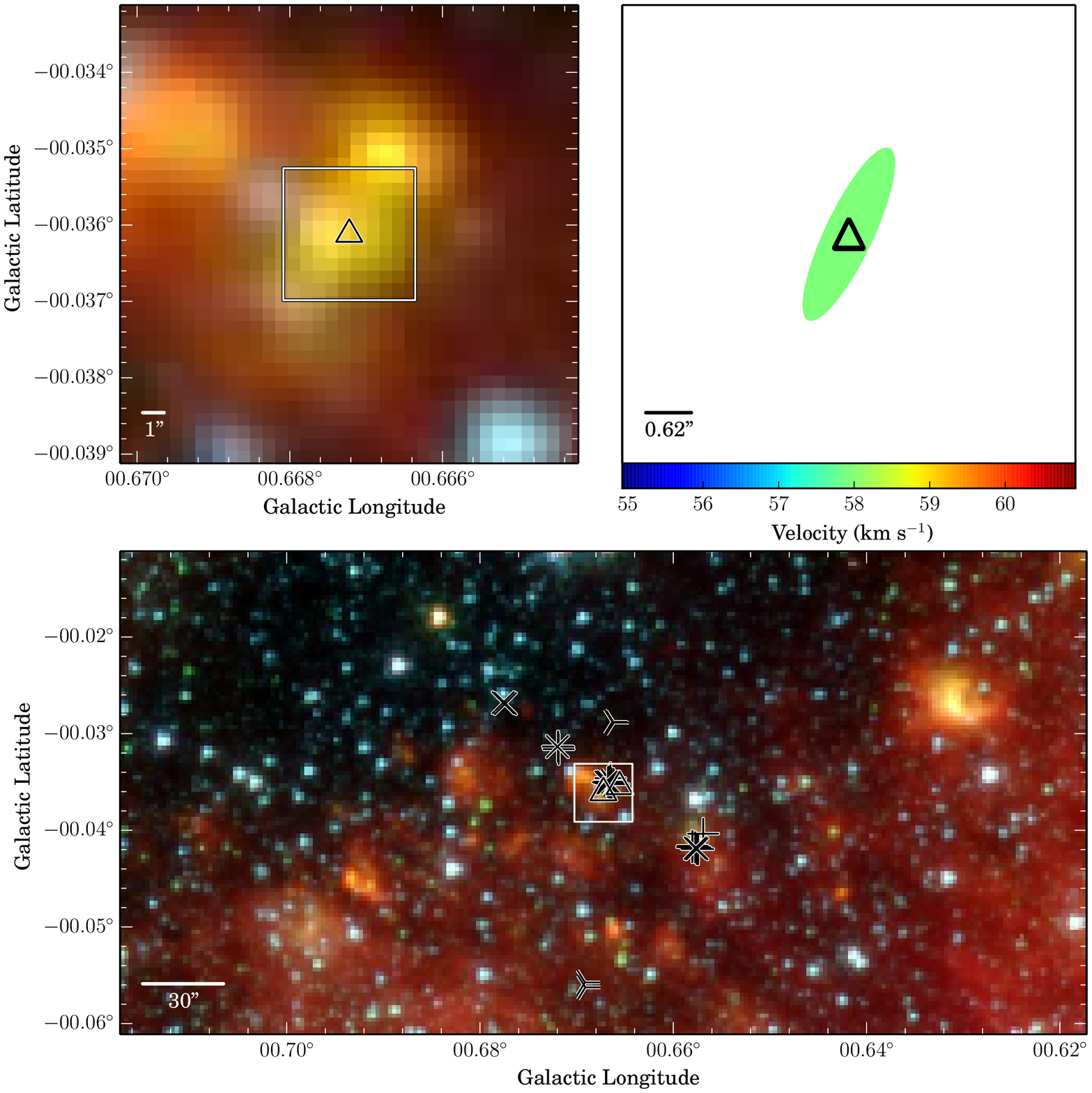}
\caption{G000.667$-$0.036 -- SF}
\label{G000.667}
\end{figure*}

\noindent{\textit{G000.669$-$0.056}}. This maser site exhibits two maser spots at 1612 MHz and has been detected by \citet{Sea1997} with a single-peaked spectrum at $+$67\kms. \citet{Are2000} and \citet{Fie2003} also detected this maser in their observations and associated this maser with the Sgr B2 star formation region. Thus, we classified this site as a star formation site. \citet{Fie2003} estimated a magnetic field of B = $-$0.7 G from a possible Zeeman Pair. In the GLIMPSE three-color image, this site is located in the extended emission background.

\noindent{\textit{G000.672$-$0.031}. This maser site is associated with a 6.7 GHz methanol maser, thus we identified it as a star formation site. This site belongs to Sgr B2N, which also shows 4.8 GHz H$_{2}$CO maser emission. This site contains three maser spots at 1665 MHz and one maser spot at 1667 MHz. The unassociated emission in the 1665 and 1667 MHz spectra is from the nearby strong star formation source G000.667$-$0.035. In the GLIMPSE three-color image, this site is located in the dark background.

\noindent{\textit{G000.678$-$0.027}}. This maser site is associated with a 6.7 GHz methanol maser, thus was identified as a star formation site. It has also been detected by \citet{Are2000} and also belongs to the well-known star formation region Sgr B2. \citet{Are2000} detected OH masers at 1612 and 1665 MHz. In our observations, we only obtained two maser spots at 1667 MHz. In the GLIMPSE three-color image, this site is associated with a bubble-like object, which may be a faint EGO.

\noindent{\textit{G000.739$+$0.410}}. This maser site is an OH/IR star site (\citealt{Hae1983}), which has also been detected by \citet{Sea1997} with an irregular spectrum at 1612 MHz. We detected six maser spots distributed in the velocity range of $-$40.2\kms to $-$3.9\kms at 1612 MHz. In the GLIMPSE three-color image, this site is associated with a bright star-like object.

\noindent{\textit{G000.810$-$1.959}}. This maser was also detected by \citet{Sea1997} with a single-peaked spectrum at $-$192.5\kms. We detected two maser spots peaked at $-$192.5\kms and $-$191.1\kms. No identification was found towards this maser site, thus we classified it into the unknown category. In the GLIMPSE three-color image, this source is associated with a bright star-like object.

\noindent{\textit{G000.892$+$1.342}}. In SIMBAD, this maser site is associated with a well-studied post-AGB star (IRAS 17393-2727) and may be undergoing the transformation to the PN stage. Meanwhile, this source presents bright [Ne II] emission (\citealt{Gae2007}) and is also associated with radio continuum emission (\citealt{Poe1987}), thus \citet{Use2012} confirmed this source as an OHPN. Therefore, we identified it as a PN site. \citet{SG2004}, \citet{Woe2012} and \citet{Goe2014} detected strong polarization in 1612, 1665 and 1667 MHz OH masers. \citet{Goe2016} detected the linear polarization at 1612 and 1667 MHz, and estimated a magnetic field of B = 6-24 mG from the circular polarization. A 22 GHz water maser at a velocity of $-$107.6\kms was also detected by \citet{Goe2015}. The presence of both OH and water masers may indicate that this PN is an extremely young PN (\citealt{Goe2015}). In the GLIMPSE three-color image, this PN is very bright at 8\um and seems to show a elliptical shape, as in Figure \ref{PN000.892}.

\begin{figure*}
\includegraphics[width=0.9\textwidth]{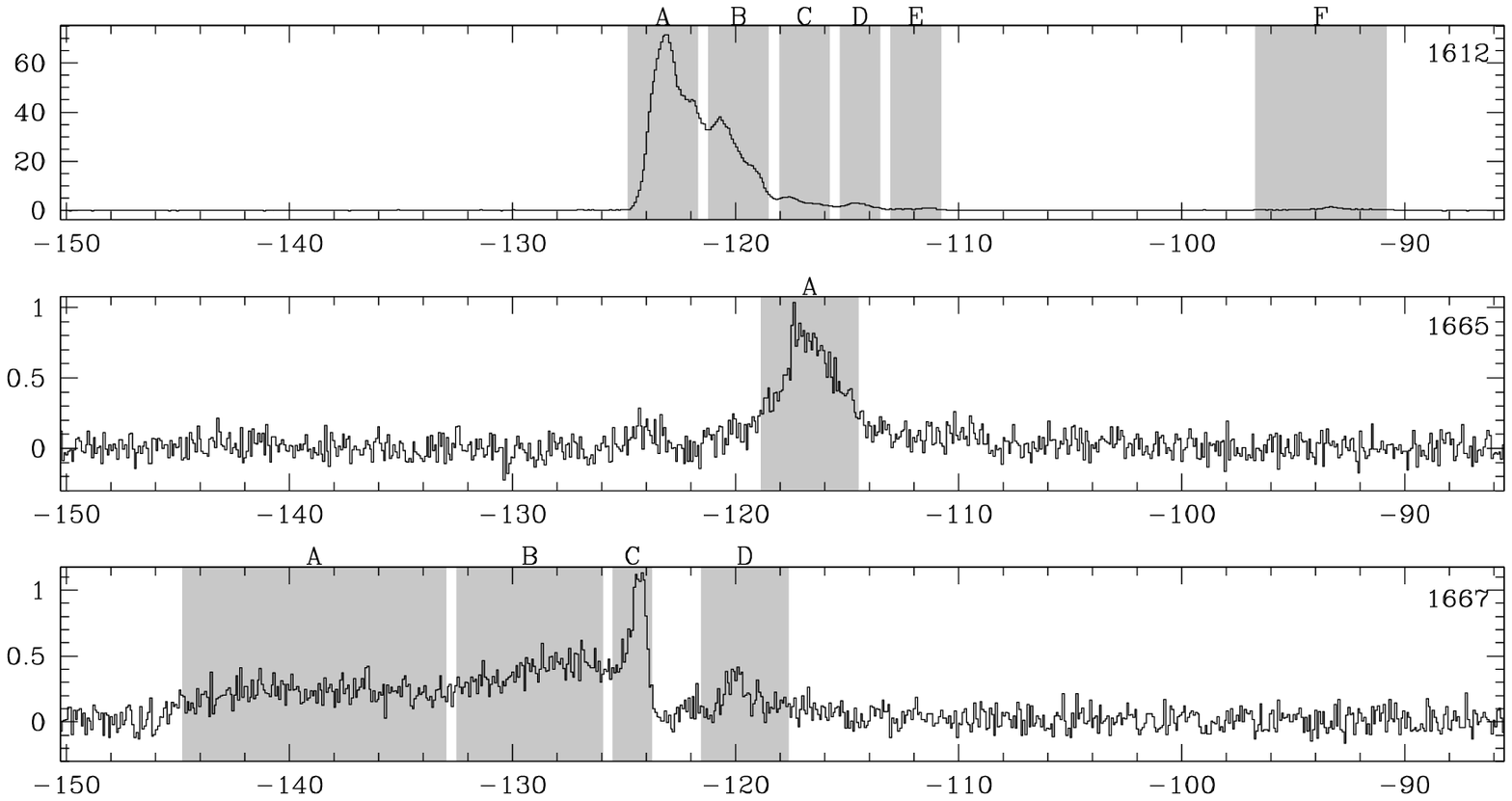}
\includegraphics[width=0.9\textwidth]{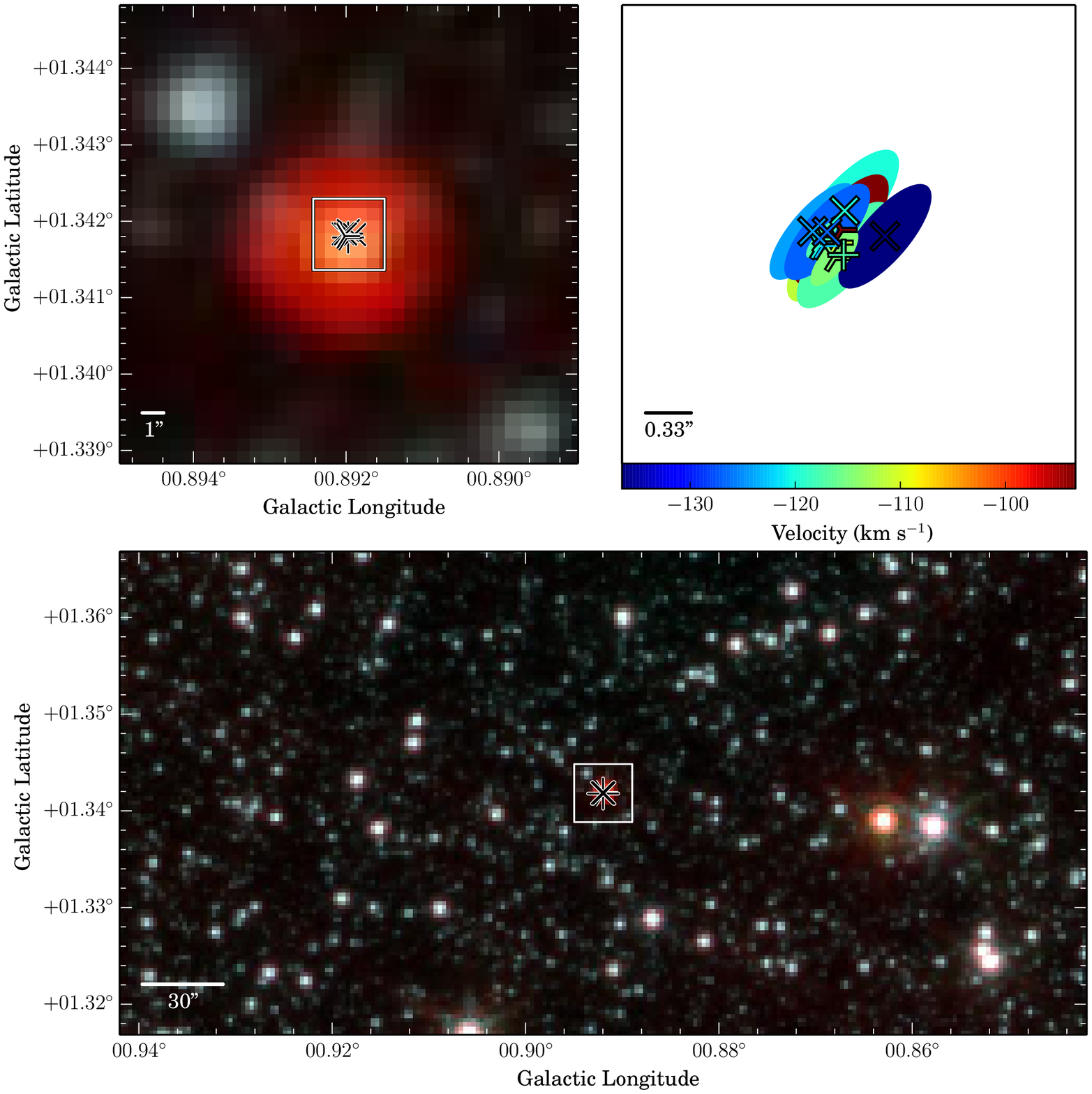}
\caption{G000.892$+$1.342 -- PN}
\label{PN000.892}
\end{figure*}

\noindent{\textit{G001.228$-$1.237}}. This maser site was identified as an evolved star site, since it shows the double-horned profile at 1612 MHz and is associated with a star-like object in the GLIMPSE three-color image. The absence of the 1667 MHz spectrum at velocities higher than $+$93.5\kms is due to the setup of the zoom bands described in Section \ref{observation}. 

\noindent{\textit{G001.240$+$0.224}}. We only detect one 1612 MHz maser spot (peaked at $-$13.1\kms) towards this maser site (IRAS 17445-2744). We searched the literature and found that this source could be listed in \citet{Tee1989}, which detected a double-peaked profile (peaked at $-$12\kms and $+$19\kms). The paper of \citet{Tee1989} is a compilation of stellar 1612 MHz maser sources in the literature. We checked the original paper \citet{Ole1984}, but it did not provide any position and velocity information. Since the observations of \citet{Ole1984} were carried out with a single-dish telescope, the position and spectrum might not be very accurate. Given the information above, we classify this maser site into the new detection category. No clear identification was found towards this source, thus we include it into the unknown category. In the GLIMPSE three-color image, this source is associated with a bright star-like object.

\noindent{\textit{G001.369$+$1.003}}. This maser site (associated with IRAS 17418-2713) is an evolved star site (\citealt{Sea1997}). We re-detected the 1612 MHz maser peaks reported by \citet{Sea1997}. Further, the 1612 MHz spectrum shows six maser spots and the 1667 MHz spectrum exhibits five maser spots. According to \citet{Gae2007} and \citet{Use2012}, this star is an AGB star, since its IR spectrum shows strong amorphous silicate absorption features and is highly variable. In the GLIMPSE three-color image, this site is associated with a very red star-like object (bright at 8\um).

\noindent{\textit{G001.671$-$0.282}}. This maser site is a new detection and only shows two 1667 MHz OH maser spots with the typical double-horned profile. In the GLIMPSE three-color image, the associated object of this maser is very bright at 8\um. We searched the literature and found no clear identification. Thus, we classified this source as an unknown maser site.

\noindent{\textit{G001.972$-$1.679}}. This maser site (IRAS 17536-2805) is associated with a bright star-like object in the GLIMPSE three-color image. \citet{Kie2004} measured a variation of about 1 magnitude in the K and J bands between two epochs, which suggests that this star is an AGB star. Thus, we include this source in the evolved star category. We only detect one maser spot at 1612 MHz.

\noindent{\textit{G002.076$+$1.738}}. This maser site (IRAS 17406-2614) is associated with a Mira type long period variable Palomar 6 V1 (\citealt{Sle2010}). Thus, we include it in the evolved star category. We only detect one maser spot at 1612 MHz. \citet{Mae2005b} detected a SiO maser towards this site with the Nobeyama telescope. In the GLIMPSE three-color image, this site is associated with a bright star-like object. 

\noindent{\textit{G002.186$-$1.660}}. This maser site is associated with an OH/IR star (IRAS 17540-2753) and was detected with a double-peaked profile at 1612 MHz by \citet{Sea1997}. In our observations, we detected six maser spots at 1612 MHz and two maser spots at 1667 MHz. Two 1667 MHz maser spots are located at similar velocities to the two strongest 1612 MHz maser spots. This source fulfils the IR color criteria for a post-AGB star (\citealt{Sue2006}). Its stable IR brightness further supports its post-AGB nature (\citealt{Rae2012}). In the GLIMPSE three-color image, the star-like object is very red (bright at 8\um).

\noindent{\textit{G002.224$+$0.461}}. This maser site is a new detection with one maser spot at 1612 MHz and a double-horned profile at 1667 MHz. We did not find any association for this source in the literature. In the GLIMPSE three-color image, this site seems to be associated with a star-like object. Based on the spectrum and IR image, we assigned this site as an unknown maser site.

\noindent{\textit{G002.602$-$0.272}}. This maser site is a new detection with a double-horned profile at 1612 MHz. In the GLIMPSE three-color image, this site is associated with a bright star-like object. Thus, it is identified as an evolved star site. The unassociated emission peaked at about $+$19\kms is from the nearby strong source G002.583$-$0.433.

\noindent{\textit{G002.640$-$0.191}}. This maser site is a re-detection of the 1612 MHz OH maser in \citet{Sea1997}. The linewidth is very broad -- about 10\kms. No identification was found towards this source in the literature. In the GLIMPSE three-color image, this source is associated with a star-like object, which may also be an EGO. Thus, we classified this source into the unknown category.

\noindent{\textit{G002.759$-$1.116}}. This 1720 MHz OH maser is a new detection. No object was found towards this site in the literature. Thus, we included this site as an unknown maser site. In the GLIMPSE three-color image, this maser site is located in the diffuse background.

\noindent{\textit{G002.818$-$0.287}}. This maser is a new detection with a double-horned profile at 1612 MHz. We searched the literature and did not find any association for this site. Moreover, in the GLIMPSE three-color image, no star-like object is associated with this site. Therefore, we classified this source as an unknown maser site.

\noindent{\textit{G003.078$-$0.027}}. This maser site has also been detected by \citet{Sea1997} and was classified as an evolved star site (\citealt{Sea1997}). The unassociated emission in the 1612 and 1667 MHz spectra is from the nearby source G003.203$+$0.024.

\noindent{\textit{G003.098$+$1.679}}. This maser site was identified as an OH/IR site by \citet{Tee1991}. We only detected one maser spot at 1612 MHz. In the WISE three-color image, this star-like object is very bright at 12\um.

\noindent{\textit{G003.304$-$2.039}}. This 1612 MHz OH maser is a re-detection of the 1612 MHz OH maser in \citet{Sea1997} and was identified as an evolved star site by \citet{Sea1997}. \citet{Soe2013} found this site is associated with a Semi-regular variable star -- OGLE BLG-LPV-177498 with a periodicity of $\sim$85 days. In our observations, we also obtained two maser spots at 1665 MHz and two maser spots at 1667 MHz. In the WISE three-color image, this source is very bright at 12\um.

\noindent{\textit{G003.472$-$1.853}}. This maser site has been detected by \citet{Sea1997} with a single-peaked spectrum at a velocity of $+$138.9\kms. We detected two maser spots at 1612 MHz and two maser spots at 1665 MHz. The linewidths of the 1612 MHz maser spots are very broad -- about 11\kms and 28\kms. The linewidths of 1665 MHz maser spots are about 11\kms. These four maser spots are detected even on the longest baselines and are not detected in the other lines, thus they are unlikely to be diffuse OH emission. This maser site was identified as a post-AGB star based on the optical spectrum in \citet{Sue2006}. In the WISE three-color image, the star-like object is very red (bright at 12\um). 

\noindent{\textit{G003.648$-$1.754}}. This maser site was identified as an OH/IR star site by \citet{Tee1991}. \citet{Sea1997} did not detect this maser in their survey. We detected one maser spot at 1612 MHz. In the WISE three-color image, this source is associated with a very red star-like object (bright at 12\um).

\noindent{\textit{G003.958$-$0.536}}. This maser site is an unknown maser site. \citet{Sea1997} detected one maser spot at $-$13.3\kms, which was also obtained by our observations. Moreover, we also detected another three maser spots with velocities higher than $-$12\kms. We did not find any associated object for this source in the literature. In the GLIMPSE three-color image, this site is associated with a star-like object.

\noindent{\textit{G004.007$+$0.915}}. This maser site is an OH/IR site (associated with IRAS 17482-2501), which has been detected by \citet{Sea1997} with a double-peaked spectrum. We not only detected the two peaks reported by \citet{Sea1997}, but also detected two 1612 MHz maser spots peaked at velocities of $+$54.4\kms and $+$60.9\kms, as well as one 1665 MHz maser spot and one 1667 MHz maser spot. This source fulfils the color criteria of a post-AGB star (\citealt{Rae2012}). In the GLIMPSE three-color image, this site is very bright at 8\um.

\noindent{\textit{G004.017$-$1.680}}. This maser site (associated with IRAS 17582-2619) is an evolved star site (\citealt{Sea1997}). This source fulfils the color criteria of post-AGB stars and does not vary in the IR flux densities, which is consistent with post-AGB stars (\citealt{Rae2012}). We also detected the typical double-horned profile at 1612 MHz. In the WISE three-color image, the site is very bright at 12\um.

\noindent{\textit{G004.565$-$0.130}}. This maser site (associated with IRAS 17535-2504) was detected with a double-peaked spectrum at 1612 MHz by \citet{Sea1997}, and thus was identified as an evolved star site. We only detected the blue-shifted maser component. This source fulfils the color criteria of post-AGB stars and does not vary in the IR flux densities, which is consistent with post-AGB stars (\citealt{Rae2012}). In the GLIMPSE three-color image, this source is associated with a  bright star-like object.

\noindent{\textit{G005.005$+$1.877}}. This maser site is a new detection and an unknown maser site. It shows one maser spot at 1612 MHz with the linewidth of about 10\kms and five maser spots at 1667 MHz. The linewidths of the five 1667 MHz maser spots range from 4\kms to 12\kms, which is quite broad. In the WISE three-color image, this source is associated with a very bright star-like object and is also very bright at 12\um. No object was found towards this site in the literature.

\section{Discussion}
\label{discussion}

\subsection{Site Categorization}

In the Galactic Center region, we identify 269 evolved star OH maser sites (76\%, two of which are associated with PNe), 31 star formation sites (9\%), four supernova remnant sites and 52 unknown sites (15\%). Compared to the literature (e.g., \citealt{Sea1997}, \citealt{Cas1998}), about 39\% of the evolved star sites (106/269) are new detections, about 45\% of the star formation sites (14/31) are new detections and about 79\% of the unknown maser sites (41/52) are new detections. Compared with the pilot region, there are relatively more sources in the evolved star and unknown categories compared to the number of sources associated with star formation. Discussion of the occurrence of the different transitions associated with evolved stars and star formation regions is given in Section \ref{overlap}. Two OH maser sites known to be associated with PNe (\citealt{Zie1989}, \citealt{Vae2001}) have been detected by our observations. OHPNe are believed to be extremely young PNe (\citealt{Zie1989}; \citealt{Use2012}). It is possible that some of other evolved star maser sites we detected also belong to this class. A proper identification would require sensitive radio continuum images and optical/IR spectroscopy. Four maser sites (\citealt{Yue1999}, \citealt{WY2002}) are associated with supernova remnants.

Out of 269 evolved star maser sites we identified towards the Galactic Centre region, \cite{Sea1997} previously detected 135, finding that most sites originated from the Galactic bulge. \citet{Sje1998} did a deep 1612 MHz OH maser survey towards the OH/IR stars in a small part of our Galactic Center region (between Galactic longitudes of $-$0.3\degree and $+$0.3\degree and Galactic latitudes of $-$0.3\degree and $+$0.3\degree). They detected 155 double-horned profiles at 1612 MHz in observations with a rms of several mJy and a velocity resolution of about 1.5\kms. Within the same region, we detected 29 1612 MHz OH masers (including seven sources identified with the mosaic described in Section \ref{observation}) and 28 of them have previously been reported by \citet{Sje1998}. 
After the usual data reduction steps introduced in Section \ref{observation}, we found that 26 maser sources in the \citet{Sje1998} sample have flux densities that surpass the 5-$\sigma$ detection limit of our Parkes survey but we failed to identify and therefore were not included in our ATCA follow-up observations. Close inspection of the Parkes survey data, in conjunction with the additional \citet{Sje1998} detections revealed that the reason we failed to identify them is that they lie in complicated absorption regions which are much stronger than the maser lines. This has meant that we were only able to identify strong maser sources in the region close to the Galactic Center with the Parkes data. Some of these \citet{Sje1998} sources fall within the fields of other stronger 1612 MHz masers that we were able to identify, but mostly fall below the detection limit of our less sensitive ATCA observations, with the exception of the seven sources discussed in Section \ref{observation}. As introduced in Section \ref{observation}, we added the ATCA images of different pointing centers together to achieve better signal to noise ratio and identify seven more evolved star maser sites. Five of them only show one maser spot at 1612 MHz (G359.716$-$0.070,  G359.939$-$0.052, G359.971$-$0.119,  G000.074$+$0.145 and G000.141$+$0.026) and two sites show the typical double-horned profile (G359.837$+$0.030 and G000.060$-$0.018). Therefore, in total, 19 sources in the \citet{Sje1998} sample with flux densities higher than the 5-$\sigma$ detection limit of our Parkes survey were not re-detected by our ATCA data.

Compared with the pilot region (in the Galactic disk), there is more dense molecular gas in the Galactic Center region (\citealt{MS1996}), but less star formation OH masers are detected towards the Galactic Center region. This may be caused by the complex environments of the Galactic Center region, such as higher temperature, higher pressure, larger velocity dispersion and larger estimated magnetic fields compared to the Galactic disk, which may suppress the star formation process (\citealt{MS1996}).

The majority of the unknown OH maser sites (31 out of 52) show one maser spot at 1612 MHz and are associated with a bright star-like object in the GLIMPSE or WISE image and are therefore likely to also originate from the circumstellar envelopes of evolved stars. Ten solitary 1720 MHz OH maser sites (not associated with any of the other three transitions of ground-state OH) are detected in the Galactic Center region. Two of them are identified as star formation sites. Four are associated with supernova remnants. We are unable to identify the exciting source associated with the remaining four solitary 1720 MHz OH maser sites, one of which (G002.759$-$1.116) is a new detection. These unknown 1720 MHz OH masers are likely to trace the shock activity in the Galactic Center region (\citealt{Yue1996}). The detailed identification process for each solitary 1720 MHz OH maser site is described in Section \ref{individual}.

\subsection{Size of Maser Sites}
\label{size}

In order to classify OH maser spots into different maser sites and thereby determine the sizes of OH maser sites, we calculated the angular separation between each maser spot and its nearest neighbour maser spot, which could be considered the lower limit of the OH maser site size. Note that each maser spot is usually considered to arise in a single, well defined position and unresolved in our observations. Figure \ref{dis_spot} shows that this angular separation is generally smaller than 2\arcsec, as was found in our pilot region analysis (\citealt{Qie2016a}). The distribution falls off quickly with increasing angular separation, with few (4\%; 37/934) spots in the 2\arcsec\ and 6\arcsec\ range. In line with \citet{Qie2016a}, we have adopted an OH maser site size upper limit of 4\arcsec. Note that this value is the same site size as adopted for water masers detected in the HOPS (\citealt{Wae2014}). After determining an upper limit for the OH maser sites, the size of the OH maser sites can be obtained by calculating the maximum distance between OH maser spots within that site. However, in two cases (G002.136$-$1.213, G000.667$-$0.035), the maser site sizes are 4.2\arcsec\ and 5.3\arcsec, respectively. G002.136$-$1.213 is an evolved star site showing a classic double-horned spectral profile representing two weak maser spots (weaker than 0.3 Jy) at 1612 MHz, thus the positional accuracy for each maser spot is relatively low. Therefore, we consider 4.2\arcsec\ to be the approximate site size of this evolved star site. G000.667$-$0.035 belongs to the complex star formation region Sgr B2M and we adopt 5.3\arcsec\ as the size of this star formation maser site, comparable to the distribution of the water maser emission detected towards this site (e.g., \citealt{Reid88})

\begin{figure}
\includegraphics[width=0.4\textwidth]{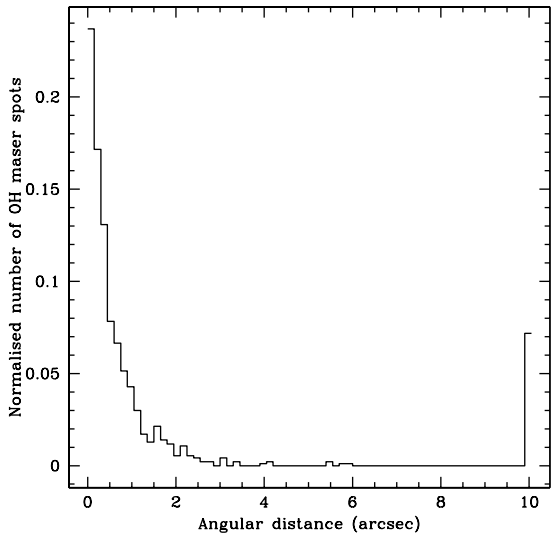}
\caption{Distribution of the angular distance to the nearest neighbour for each maser spot. This figure is cut off at 10\arcsec\ where a higher number of unrelated nearby sources start to show.}
\label{dis_spot}
\end{figure}

\begin{figure}
\includegraphics[width=0.4\textwidth]{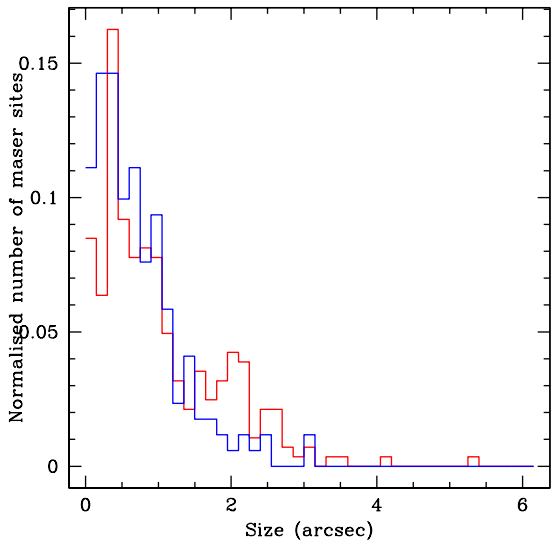}
\caption{Distribution of the sizes of 283 OH maser sites in this paper (red) and 171 OH maser sites in the SPLASH pilot region (blue) from \citet{Qie2016a}. All these maser sites exhibit more than one maser spot and include evolved star sites, star formation sites and unknown sites. Only six OH maser sites in this paper are larger than 3\arcsec\ and five of them are associated with evolved stars.}
\label{size_all}
\end{figure}

\begin{figure}
\includegraphics[width=0.4\textwidth]{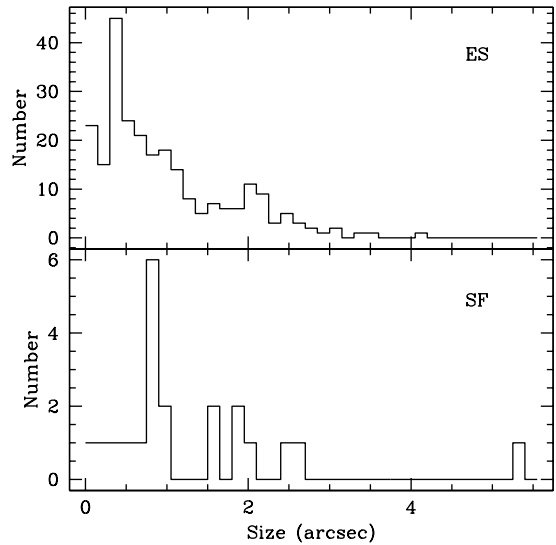}
\caption{Distribution of the sizes of 248 evolved star OH maser sites (top) and 21 star formation OH maser sites (bottom).}
\label{size_es_sf}
\end{figure}

Figure \ref{size_all} illustrates the sizes of 283 OH maser sites (red) in this paper and the sizes of 171 OH maser sites (blue) in the SPLASH pilot region. For the Galactic Center region, Figure \ref{size_all} shows that the majority of the OH maser sites (98\%; 277/283) are smaller than 3\arcsec, which is equivalent to a linear size of about 0.12 pc at a distance of 8 kpc. The fraction of OH maser sites with sizes smaller than 2\arcsec\ is 86\% (243/283), which is significantly smaller than the value (95\%) in the pilot region. A Kolmogorov-Smirnov (K-S) test (under the null hypothesis that the OH maser sites in the Galactic Center and the pilot regions are drawn from populations with the same size distribution) gave an asymptotic probability $p=4\times 10^{-3}$. Therefore, there is strong evidence that their size distribution is different. There are 34 OH maser sites in the Galactic Center region that show a site size between 2\arcsec\ and 3\arcsec\ and we find 27 of these sites arise from the circumstellar envelopes of evolved stars, three of them are from star formation regions and four are from the unknown sites. Thus, we conclude that the OH maser sites in the Galactic Center region are generally larger than the OH maser sites in the SPLASH pilot region and this may be mainly due to the larger sizes of evolved star sites in the Galactic Center region (none of the evolved star sites are distributed over more than 2\arcsec\ in the pilot region but 13\% of the evolved stars in the Galactic Center region are). 

Restricting our analysis to the evolved star sites, we compared their
size distribution in the Galactic Center region  (this paper) with
that in the pilot region (Figure 4 in \citealt{Qie2016a}). Their size
distribution is clearly different (a K-S test gives
$p=2\times10^{-4}$). In fact, we find strong evidence that the size
of OH masers in the Galactic Center region is larger than
in the pilot region. The median sizes of the samples are 0.72\arcsec\ and
0.43\arcsec, respectively. We confirmed the statistical significance
of this difference with a Mann-Whitney U test, giving a very low p-value of 
$3\times 10^{-7}$, meaning that the hypothesis that the two medians are equal can be rejected.
A possible interpretation of this size difference could be that the
evolved star OH maser sites in the pilot region are typically farther
away than those in the Galactic Center. This would be unlikely if we
assume a smooth distribution of evolved stars, with a higher number
density toward the center. For instance, for a nearly constant AGB
number density in the inner 5 kpc, and a rapid decrease at larger
radii (\citealt{Jae2002}), we would expect a source median
distance of $\simeq 8$ kpc in the Galactic Center region, and $< 9$
kpc in the pilot region. This cannot explain a ratio of median sizes
of $\simeq 1.7$. However, we cannot rule out other biases in the
typical distances of both samples, due to the Galactic spiral
structure. A more detailed analysis, with a determination of distances
to the maser sites would be necessary to study any possible bias.

Alternative explanations could be that the maser sites in the
Galactic Center are intrinsically larger, due to higher expansion
velocities in their circumstellar envelopes or older ages of the
evolved stars. Higher expansion velocities are certainly possible,
given that the metallicity of stars towards the Galactic Center is
higher (\citealt{FC2013}), yielding larger expansion rates in
the AGB phase (e.g., \citealt{Goe2017}). This would naturally lead
to larger linear sizes. However, we see no evidence in our data that
the expansion velocity of the evolved stars (derived from the velocity
difference between the red- and blue-shifted OH maser peaks) in the
Galactic Center sample is higher than those in the pilot region. A
possible difference in typical age is still an open possibility, if
the sources in the Galactic Center region are in a more advanced stage
of stellar evolution.

We also studied the size distributions of 248 evolved star sites and 21 star formation sites, shown in Figure \ref{size_es_sf}. Similarly to \citet{Qie2016a}, we find that the star formation OH maser sites peak at a larger angular size than the evolved star OH maser sites. As discussed in \citet{Qie2016a}, this is not surprising, since the star formation OH masers are distributed over the compact \hii region about 3000 AU (\citealt{FC1989}), whereas the evolved star OH masers trace the circumstellar envelopes, typically on scales of about 80 AU (\citealt{Rei2002}). The site size of star formation masers may be affected by the small number statistics. 


\subsection{Overlap between OH Transitions}
\label{overlap}
In our observations, many OH maser sites only show maser emission in one OH transition. However, some maser sites exhibit more than one transition. The upper panel of Figure \ref{vennfig} shows the overlap between the four ground-state OH transitions towards evolved star sites. Of the 269 evolved star maser sites, 226 sites (84\%) only show the 1612 MHz maser emission, a similar percentage to that found in the SPLASH pilot region (83\%). Among these 226 evolved star maser sites showing just 1612 MHz OH emission, 201 maser sites show double-horned spectral profiles in the 1612 MHz spectra, 21 maser sites only exhibit one maser spot at 1612 MHz, one site (G359.260$+$0.164) has two maser spots, one site (G359.150$-$0.043) shows three maser spots, one site (G356.457$-$0.386) exhibits four maser spots and the remaining one (G000.739$+$0.410) presents six maser spots. One evolved star site (G358.656$-$1.710) only exhibits mainline transitions with one maser spot at 1665 MHz and one maser spot at 1667 MHz. Note that, in the evolved star category, only this site (G358.656$-$1.710) did not show the 1612 MHz OH maser emission. No evolved star sites show the 1720 MHz OH emission. 1612 and 1667 MHz OH masers have the largest overlap: 97\% of the 1667 MHz OH maser sites (36/37) also show 1612 OH maser emission. Unlike \citet{Qie2016a} who found that the mainline transitions had the second largest overlap, in the Galactic Center region we find that 1612 and 1665 MHz OH masers show the second largest overlap with 94\% of the 1665 MHz OH masers (16/17) having a 1612 MHz OH maser counterpart.

\begin{figure}
\includegraphics[width=0.4\textwidth]{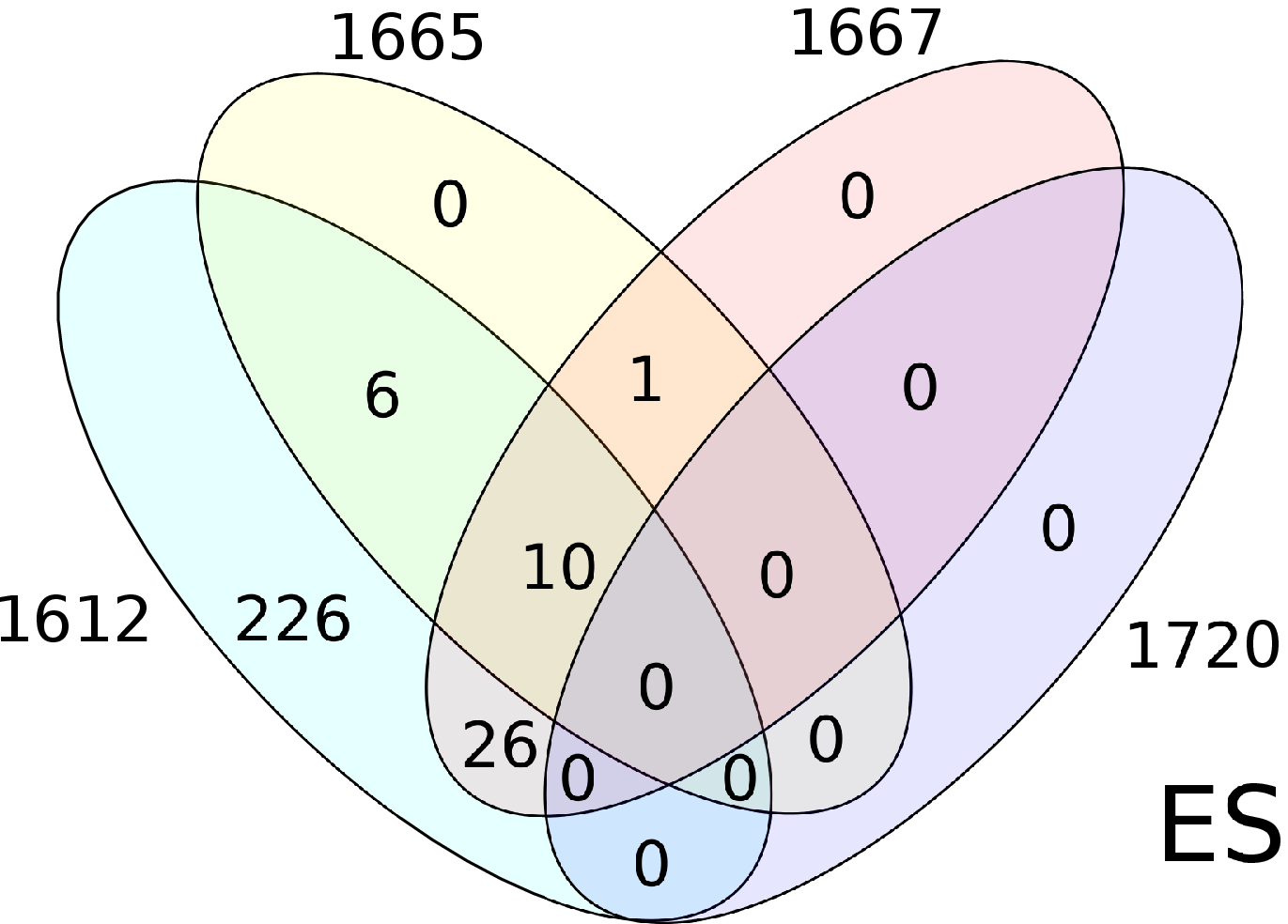}
\includegraphics[width=0.4\textwidth]{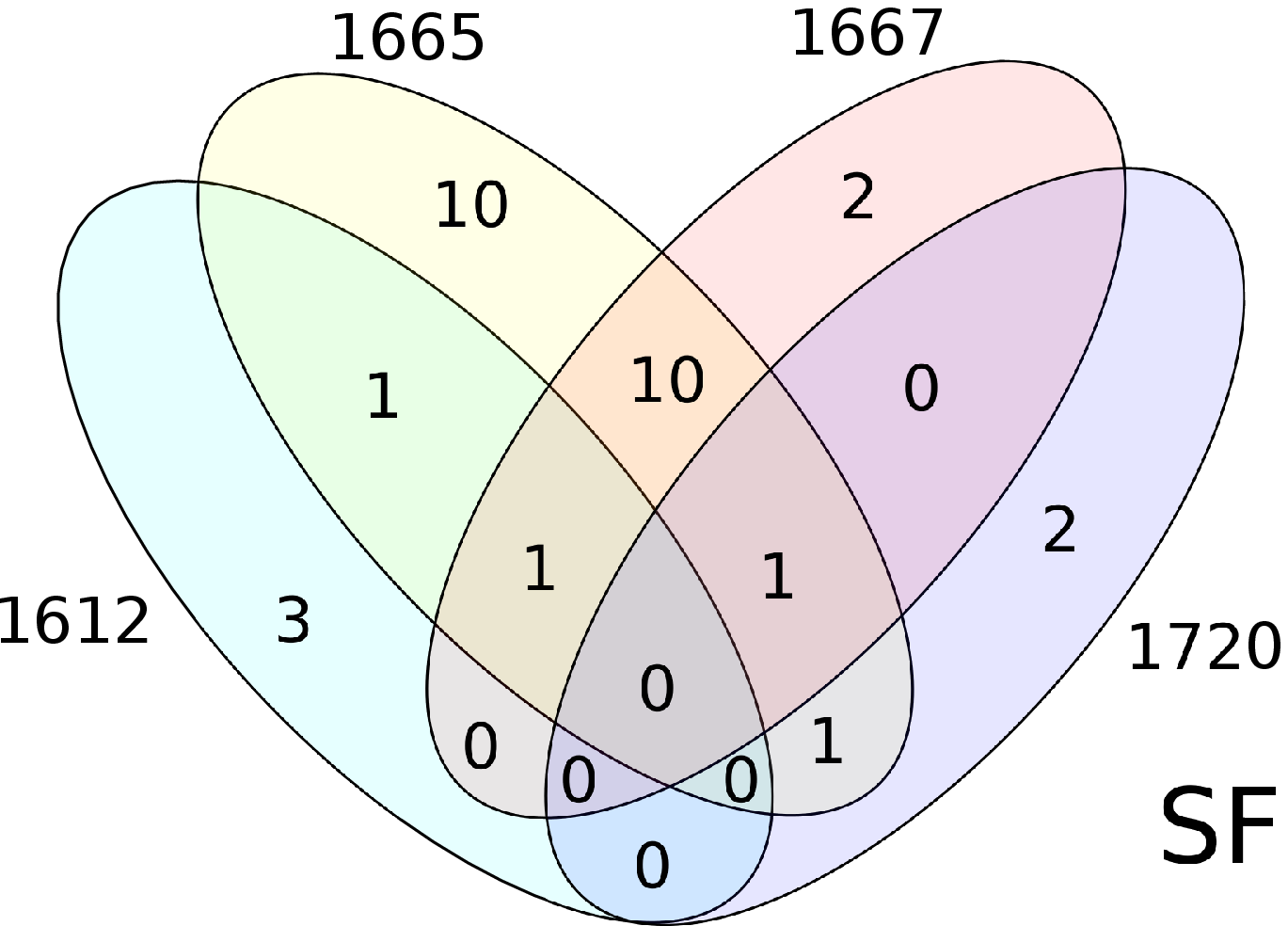}
\caption{Upper panel is a Venn diagram, which shows the transition overlap of evolved star (ES) OH maser sites. Bottom panel is a Venn diagram, which describes the transition overlap of star formation (SF) OH maser sites.}
\label{vennfig}
\end{figure}

The bottom panel of Figure \ref{vennfig} shows the transition overlap of the star formation category. Since there are 31 star formation maser sites in the Galactic Center region, this result may be affected by the small number statistics. The majority of star formation maser sites (24/31) show the 1665 MHz OH maser emission and 60\% of 1612 MHz (3/5) and 50\% of 1720 MHz (2/4) OH masers are solitary, i.e., not associated with any of the other three ground-state OH transitions. The fraction of solitary 1665 MHz OH masers (10/24) is about 42\%, which is higher than that in the pilot region (33\%). Mainline transitions show the largest overlap: 50\% of the 1665 MHz OH masers (12/24) are associated with the 1667 MHz OH maser emission. This value is lower than the fraction (62\%) reported in \citet{Qie2016a} for the SPLASH pilot region. 

\citet{Cas1998} detected 16 star formation OH maser sites in the Galactic Center region and we re-detected 12 of them. Among the four non-detections, three of them were very weak at 1665 MHz (weaker than 0.4 Jy) and one of them was about 0.7 Jy at 1665 MHz. \citet{Cae2013} used the Parkes telescope to obtain full polarisation spectra for all 1665 and 1667 MHz OH masers accessible to the Parkes telescope. They re-detected these four non-detections, all of which are weaker than 0.4 Jy at 1665 MHz. Given that the typical 5$\sigma$ detection limit of our ATCA observations is about 0.4 Jy, we suggest that three of the four non-detections are due to the lower sensitivity of our observations. The remaining one (with a flux density of about 0.7 Jy) is due to the variability of the OH maser emission, since the time between our observations and those of \citet{Cas1998} is about 20 years. For the re-detected 12 OH masers, in \citet{Cas1998}, 11 (out of 12) sites exhibit both 1665 and 1667 MHz OH masers, however, in our observations, eight OH masers show both mainline transitions. Three sources with both mainline transitions in \citet{Cas1998} were only detected with the 1665 or 1667 MHz transition in our observations. For these three sources, \citet{Cas1998} reported the peak flux densities of 0.4 (at 1667 MHz), 0.4 (at 1667 MHz) and 1.6 Jy (at 1665 MHz). In \citet{Cae2013}, they re-detected these three sources with the peak flux densities of 0.7 (at 1667 MHz), 0.8 (at 1667 MHz) and 1.2 Jy (at 1665 MHz), respectively. Thus, we attribute our non-detections to temporal variability in the time between the observations. Further studies on the full SPLASH region will allow us to precisely determine the transition overlap for both evolved star and star formation OH maser categories.    

%
%

\subsection{Evolved Star Sites}
\label{evolvedstar}

For 266 evolved star maser sites (excluding two PN sites and one evolved star site only showing mainline transitions) with 1612 MHz OH maser emission, we categorize them based on the ratio of the integrated flux densities of their blue-shifted and red-shifted components (I$_{blue}$ and I$_{red}$). As described in \citet{Qie2016a}, we can use the integrated flux densities to estimate the number of photons from each side of the circumstellar envelope of the evolved star. We adopt the same criterion as the pilot region paper. If the ratio of I$_{blue}$ and I$_{red}$ is between 0.5 and 2, we classify the source into the symmetric category. Otherwise, we include the evolved star site in the asymmetric category, including 24 sources that show one maser spot at 1612 MHz (after ensuring that the ratio was satisfied using the integrated flux density of the detected feature and the 5-$\sigma$ detection limit).

\begin{figure}
\includegraphics[width=0.4\textwidth]{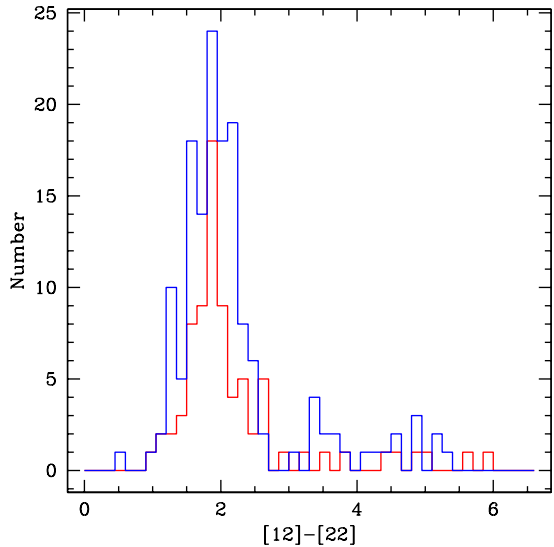}
\caption{WISE [12]-[22] color distributions of 149 symmetric (blue) and 78 asymmetric (red) evolved star sites showing 1612 MHz transition. Note that 28 symmetric sites and 11 asymmetric sites do not have the WISE [12]-[22] color, thus are not included here.}
\label{hist_1222}
\end{figure}

\begin{figure}
\includegraphics[width=0.4\textwidth]{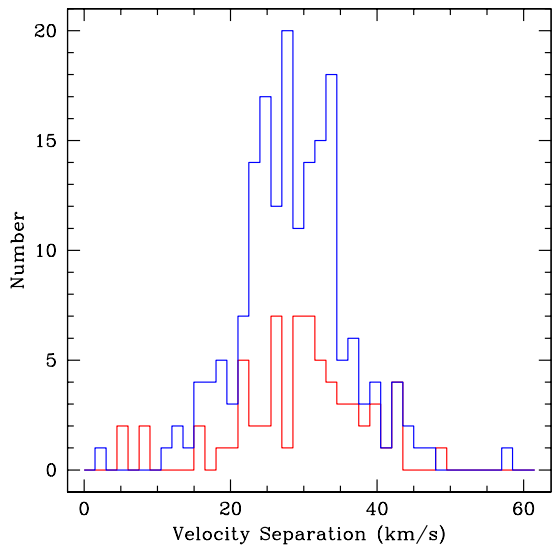}
\caption{The velocity separation $\bigtriangleup V$ of 177 symmetric (blue) and 65 asymmetric (red) evolved star sites showing 1612 MHz transition. Note that 24 asymmetric sources with only one maser spot at 1612 MHz are not included here.}
\label{Vel_Sep}
\end{figure}

We find that there are 177 (out of 266, 67\%) symmetric sources and 89 (out of 266, 33\%) asymmetric sources. Comparison between the mid-IR properties of these two samples is made, since for wavelengths longer than 5\um, emission from the circumstellar dust can become the dominant emission source compared with the radiation from the photosphere (\citealt{Ble2006}). We are able to calculate the WISE [12]-[22] color for 149 (84\%) symmetric sources and 78 (88\%) asymmetric sources. Unlike \citet{Qie2016a}, we did not find obvious difference in the WISE [12]-[22] color (12\um and 22\um) of these two samples, shown in Figure \ref{hist_1222}. A K-S test yields $p=0.94$, so it is consistent with both distributions being drawn from the same underlying population. Figure \ref{Vel_Sep} shows the velocity separation of the most extreme spectral features associated with symmetric and asymmetric sources (excluding the 24 sources exhibiting single spectral features). While the plot shows no clear offset between the two categories, a K-S test shows $p=0.05$, so there is marginal evidence that the two distributions are not drawn from the same underlying population. Thus, these results seem to suggest that symmetric and asymmetric evolved star OH maser sites in the Galactic Center region exhibit similar mid-IR color properties and velocity separation properties. There could be some biases that should be considered, e.g., the particular ratio of I$_{blue}$ and I$_{red}$ we chose to classify these two samples. The detailed studies on the longitude-velocity distribution of the evolved star masers are ongoing in a related paper (Imai et al. in preparation). Further investigation will be conducted with the full SPLASH region.

\subsection{Star Formation Sites}
\label{starformation}

In the Galactic Center region, we detected 31 star formation OH maser sites. We have compared the occurrence of the star formation OH maser sites that we detect with 6.7 GHz methanol masers detected in the MMB survey (\citealt{Cae2010}) and water masers from HOPS (\citealt{Wae2014}) in the overlapping survey region of Galactic longitudes of $355^{\circ}$ to $5^{\circ}$ and Galactic latitudes of $-0.5^{\circ}$ to $+0.5^{\circ}$. 
We used an association threshold of 5.1\arcsec\ (corresponding to a 0.001\degree in both Galactic longitude and latitude) to determine associations between OH, methanol and water masers. There are 27 OH masers located in this region and 20 (74\%) show 6.7 GHz methanol masers, comparable to the 73\% association rate found in the pilot region (\citealt{Qie2016a}). Ten OH maser sites (37\%) exhibit 22 GHz water masers and this fraction is comparable to the 44\% association rate found in the pilot region (\citealt{Qie2016a}). Nine of these 27 OH maser sites have both 6.7 GHz methanol masers and 22 GHz water masers. Moreover, as discussed in \citet{Qie2016a}, these results between OH/methanol and water masers are affected by the low sensitivity of HOPS, which has a typical rms noise of about 1 Jy. Thus, we checked \citet{Bre2010} and \citet{Tie2016}, which are sensitive 22 GHz water maser surveys targeted towards the OH and MMB masers (a typical rms noise of about 0.1 Jy). After comparing these OH masers with water masers from \citet{Bre2010}, \citet{Tie2016} and \citet{Wae2014}, we found that 16 OH masers (out of 27; 59\%) are associated with water masers. Further investigation with the full SPLASH region will be conducted. 

We investigated the 1.7 GHz radio continuum properties (utilising the 1720 MHz zoom bands) towards each of these 31 OH maser sites and find that six of them (19\%) are associated with radio continuum sources. This detection rate is higher than the 9\% found in the pilot region, but still lower than the 38\% association rate found by \cite{FC2000} for a sample of OH masers and radio continuum sources detected at 8.2 and 9.2 GHz (a typical rms noise of about 0.15 mJy). This result may be affected by some biases, e.g., nine (out of 31) star formation OH maser sites belong to the complex star formation region Sgr B2 and five of these nine star formation sites show continuum emission at 1.7 GHz. Small number statistics can also cause some biases. As discussed in \citet{Qie2016a}, the low continuum association rate that we find may be caused by the low frequency (1.7 GHz) and limited sensitivity of our continuum observations (a typical rms noise of about 10 mJy). 

\subsection{Nondetection Sources}
\label{nodetections}

\tabletypesize{\small}
\begin{table}
\begin{center}
\caption{\textnormal{List of positions which exhibit maser emission in Parkes observations, but were not detected in our ATCA observations.}}
\label{nondetection}
\begin{tabular}{cc}
\hline
\hline
G355.30$+$1.80(D,W)&G356.30$-$1.65(D,W)\\
G356.55$+$0.85(D,Z)&G356.55$-$1.00(D,W)\\
G356.65$+$0.10(D,Z)&G357.40$+$1.225(A,W)\\
G358.65$+$1.60(D,W)&G358.90$+$1.55(A,S)\\
G002.15$+$0.80(A,W)&\\

\hline
\end{tabular}
\end{center}

\textbf{Notes.} D -- a double-horned spectrum; A -- a single-peaked spectrum; W -- weak ($<$0.3 Jy) in Parkes observations; Z -- due to the coverage of zoom bands; S -- spurious detection in the Parkes observations. 


\end{table}

Table \ref{nondetection} presents the positions of OH maser candidates from the Parkes observations towards which we did not detect any maser emission in our ATCA observations. Two of them (G356.55$+$0.85 and G356.65$+$0.10) were not detected due to the setup of zoom bands, which did not cover the velocity range of these two maser sites. One site (G358.90$+$1.55) was very weak ($\sim$2$\sigma$) at 1667 MHz in the Parkes spectrum and we therefore consider the original Parkes identification as spurious. The remaining six sources appear to be real, but weak detections (weaker than 0.3 Jy) in the Parkes observations: four of them showed the typical double-horned profile in the 1612 MHz (3/4) or 1667 MHz (1/4) spectra and two of them only exhibited one peak in the 1612 MHz spectra. Given that the typical 5$\sigma$ detection limit of our ATCA observations is 0.4 Jy, it is entirely possible that our follow-up observations simply had inadequate sensitivity to detect the six weak sources, especially if they had reduced in peak flux density in the 3 years between the Parkes and ATCA observations. Small levels of temporal variability are common in ground-state OH masers, in fact \cite{Cae2014} found that less than 10\% of a sample of 187 OH masers remained stable over a period of decades.
   
\section{Conclusions}

In this paper, we report accurate positions for ground-state OH masers in the SPLASH Galactic Center region, between Galactic longitudes of $355^{\circ}$ and $5^{\circ}$ and Galactic latitudes of $-2^{\circ}$ and $+2^{\circ}$. We detect a total of 356 maser sites, which show maser emission in one, two or three transitions. About half of these 356 maser sites (161/356) have been newly discovered by the SPLASH observations. We also identify the associated astrophysical objects for these maser sites. 

269 OH maser sites are associated with evolved stars (including two PN sites). These maser sites usually exhibit the typical double-horned profile at 1612 MHz, occasionally accompanied by 1665 and/or 1667 MHz OH masers. 31 maser sites are classified as star formation sites and commonly show several strong maser spots in mainline transitions and occasionally also exhibit 1612 or 1720 MHz OH masers. Four maser sites are associated with supernova remnants. 52 maser sites are categorised as unknown maser sites due to the lack of complementary information from the literature and IR images. 

We find the size of most OH maser sites (98\%) are smaller than 3$\arcsec$ based on their accurate positions. Compared with the OH maser sites in the pilot region, the OH maser sites in the Galactic Center region generally have larger angular sizes. In the absence of evidence for differences in expansion velocity, we suggest that this may possibly be due to older characteristic ages (and hence larger linear sizes) of the Galactic Center evolved star population compared to the pilot region evolved star population.

We categorize evolved star sites based on the integrated flux densities of blue- and red-shifted components at 1612 MHz, and, unlike the pilot region paper, we find no obvious difference in the WISE [12]-[22] colors of the symmetric sources and asymmetric sources. 

We find that six of the 31 star formation OH maser sites in the SPLASH Galactic Center region are associated with continuum sources at 1.7 GHz, which is higher than the ratio in the pilot region and lower than the ratio of OH maser sites associated with 9 GHz (\citealt{FC2000}). This result is likely due to several biases, e.g., small number statistics and the close evolutionary stages of sources in the Sgr B2 region. The frequency we observe and the low sensitivity of our continuum observations may also play an important role.

We did not detect any maser emission in nine target fields. From their Parkes spectra, we find that they tend to show simpler and weaker profiles. We consider two thirds of the non-detections are likely due to intrinsic variability.

\acknowledgments The Australia Telescope Compact Array is part of the Australia Telescope which is funded by the Commonwealth of Australia for operation as a National Facility managed by CSIRO. This research has made use of: NASA's Astrophysics Data System Abstract Service; and the SIMBAD data base, operated at CDS, Strasbourg, France. This work is based in part on observations made with the Spitzer Space Telescope, which is operated by the Jet Propulsion Laboratory, California Institute of Technology under a contract with National Aeronautics and Space Administration (NASA). This publication also makes use of data products from the Wide-field Infrared Survey Explorer, which is a joint project of the University of California Institute of Technology, funded by NASA. H.-H.Q. is partially supported by the Special Funding for Advanced Users, budgeted and administrated by Center for Astronomical Mega-Science, Chinese Academy of Sciences (CAMS-CAS) and CAS ``Light of West China'' Program. JFG is partially supported by MINECO (Spain) grants AYA2014-57369-C3-3 
and AYA2017-84390-C2-1-R (co-funded by FEDER). This work was supported in part by the Major Program of the National Natural Science Foundation of China (Grant No. 11590780, 11590784) and the CAS Pioneer Hundred Talents Program(Technological excellence, Y650YC1201).

\clearpage


\label{lastpage}

\end{document}